\documentclass[preprint,aps,showpacs,nofootinbib,preprintnumbers,amsmath,amssymb,superscriptaddress]{revtex4-2}
\usepackage{amsmath,amssymb,amsthm,mathrsfs}
\usepackage{epsfig}
\usepackage{graphicx}
\usepackage[usenames,dvipsnames]{color}
\usepackage{subfigure}
\usepackage{slashed}
\usepackage{comment}
\usepackage[normalem]{ulem}
\usepackage[compat=1.1.0]{tikz-feynman}
\usepackage[colorlinks,citecolor=blue]{hyperref}
\usepackage{color}
\usepackage{hyperref}
\usepackage{multirow}
\tikzfeynmanset{compat=1.1.0}

\usepackage{xcolor}
\usepackage{soul}

\begin{document}

\title{Cosmological Constraints on Temperature-Dependent Interaction between Dark Matter and Neutrinos }
\author{Ren-Peng Zhou\footnote{zhourenpeng21@mails.ucas.ac.cn}}
\affiliation{School of Fundamental Physics and Mathematical Sciences, Hangzhou Institute for Advanced Study, UCAS, Hangzhou 310024, China}
\affiliation{University of Chinese Academy of Sciences (UCAS), Beijing 100049, China}
\affiliation{Institute of Theoretical Physics, Chinese Academy of Sciences, Beijing 100190, China}
\author{Da Huang\footnote{Corresponding author: dahuang@bao.ac.cn}}
\affiliation{National Astronomical Observatories, CAS, Beijing, 100012, China}
\affiliation{School of Fundamental Physics and Mathematical Sciences, Hangzhou Institute for Advanced Study, UCAS, Hangzhou 310024, China}
\affiliation{University of Chinese Academy of Sciences (UCAS), Beijing 100049, China}
\affiliation{International Centre for Theoretical Physics Asia-Pacific, Beijing/Hangzhou, China}
\begin{abstract}
\noindent We study the influence of the temperature-dependent interaction between dark matter (DM) and neutrinos on the measurement of cosmological parameters. We pay attention to the neutrino mass effects, so that the derivation of Boltzmann equations needs to specify the concrete form of interaction. We work in a model in which the DM-neutrino scatterings are induced by a dimension-six operator, and present the details for deriving the full Boltzmann hierarchy for DM and neutrinos, including a novel method to obtain the fluid approximation for modes entering the horizon. It is shown that our interaction can induce the dark acoustic oscillation in the DM-neutrino fluid, leaving distinct signatures on the CMB and matter power spectra. By using the latest CMB and BAO datasets from Planck, DESI and ACT, the constraint on today's DM-neutrino interaction parameter for the normal neutrino mass ordering reaches $u^0_{\chi-\nu} \lesssim {\cal O}(10^{-13})$, nearly nine orders stronger than that for temperature-independent case in the literature. This can be understood by noting that the scattering cross section increases nearly quadratically with cosmological temperature in the early universe, leading to enhanced effects. We have investigated alternative scenarios with different neutrino mass assumptions. In particular, models with degenerate neutrino masses give rise to weaker constraint of $u^0_{\chi-\nu} \lesssim {\cal O}(10^{-11})$, showing the importance to incorporate the realistic neutrino mass ordering in the fits. Finally, when employing the logarithmic flat prior for $u^0_{\chi-\nu}$, we have shown hints to a nonzero interaction at $95\%$ CL by combining Planck, DESI and ACT data.  

\end{abstract}

\maketitle
\newpage
\section{introduction}
\noindent The present standard cosmological  $\Lambda$CDM model has achieved a high level of success across a wide range of scales. It is able to simultaneously account for the Cosmic Microwave Background (CMB) anisotropies and the formation of Large-Scale Structure (LSS) in the universe. In this model, dark matter (DM) is assumed to be cold and collisionless. This is consistent with the present observations from terrestrial DM detection experiments, which have not yet yielded definitive evidences of interactions between DM and Standard Model (SM) particles~\cite{Roszkowski:2017nbc,Arcadi:2017kky,Boveia:2018yeb,Schumann:2019eaa,Cirelli:2024ssz,Liang:2024ecw,Liang:2025kkl}. 
However, searches for signals of the non-gravitational DM interactions remains crucial, as it could reveal the origin and properties of DM~\cite{Roszkowski:2017nbc,Boveia:2022adi,Arcadi:2024ukq}.

In search for interactions between dark and visible sectors, cosmological surveys play an essential role. Since such interactions can leave imprints on the CMB and matter power spectra, the associated cosmological data can be used to constrain their strength. In recent years, a series of extensive studies have been conducted on the DM interactions with 
baryons~\cite{Dvorkin:2013cea,Munoz:2015bca,Slatyer:2018aqg,Xu:2018efh,Stuart:2024nkz,He:2025npy,Dhyani:2025uof,Rahimieh:2025fsb,Gluscevic:2017ywp,Boddy:2018kfv,Boddy:2018wzy,Dvorkin:2020xga,Nguyen:2021cnb,Boddy:2022tyt,Ali-Haimoud:2023pbi,Barkana:2018lgd,Munoz:2017qpy,Ali-Haimoud:2015pwa}, photons~\cite{Wilkinson:2013kia, Stadler:2018jin, Becker:2020hzj, Weiner:2012cb, Boehm:2014vja, Kumar:2018yhh, Escudero:2018thh,Zhou:2022ygn,Yadav:2019jio}, and neutrinos~\cite{Boehm:2000gq,Boehm:2004th,Wilkinson:2014ksa,Serra:2009uu,Shoemaker:2013tda,Boehm:2014vja,Bertoni:2014mva,Escudero:2015yka,DiValentino:2017oaw,Olivares-DelCampo:2017feq,Escudero:2018thh,Becker:2020hzj,Mangano:2006mp,Mosbech:2020ahp,Dev:2025tdv,Trojanowski:2025oro,Zapata:2025huq,Zu:2025lrk,Mosbech:2024wxr,Heston:2024ljf,Giare:2023qqn,Dey:2023sxx,Pal:2023dcs,Akita:2023yga,Brax:2023tvn,Brax:2023rrf,Cline:2023tkp,Cline:2022qld,Dey:2022ini,Hooper:2021rjc,Paul:2021ewd,Crumrine:2024sdn,Bertolez-Martinez:2025trs,Stadler:2019dii,Diacoumis:2018ezi,Oldengott:2014qra,Bringmann:2013vra,Audren:2014lsa,Horiuchi:2015qri,Pandey:2018wvh,Choi:2019ixb}. In particular, recent analyses of data from Lyman-$\alpha$~\cite{Hooper:2021rjc}, CMB~\cite{Brax:2023rrf,Brax:2023tvn,Giare:2023qqn}, 
and weak lensing~\cite{Zu:2025lrk} have suggested a potentially non-negligible DM-neutrino interaction. The existence of such a coupling can affect the cosmological history and generate the imprints on the CMB and matter power spectrum. 
Concretely, with its scattering to neutrinos, DM could contribute to the fast-evolving mode, thereby influencing the growth of baryon-photon perturbations~\cite{Wilkinson:2014ksa}. The free-streaming of neutrinos is also altered, giving rise to the reduced anisotropic stress and the strengthened clustering, which in turn enhances the gravitational boost effect~\cite{Mangano:2006mp}. 
Moreover, the pressure from neutrinos can compete with the gravitational attraction among DM particles so as to induce the diffusion-damped dark acoustic oscillations (DAO), which would become evident in the matter power and CMB spectra at small scales. 

Note that the cosmological investigation of DM-neutrino interactions has already done in numerous earlier studies under diverse theoretical assumptions and simplifications. 
For example, Ref.~\cite{Wilkinson:2014ksa} assumed neutrinos to be massless and considered the scenarios with a constant or $T^2$-dependent DM-neutrino scattering cross section, where $T$ represented the neutrino temperature. In Ref.~\cite{Mosbech:2020ahp}, the authors studied the neutrino mass effect by carefully deriving the full Boltzmann hierarchy in the case with temperature-independent DM-neutrino interactions. One presumption in the latter work was that the three neutrinos were degenerate in mass, which is obviously disfavored in view of the current neutrino oscillation data~\cite{ParticleDataGroup:2024cfk}. It is clear that the previous constraints on DM-neutrino interactions relied on certain assumptions in the neutrino masses and interaction forms. In order to fully explore the DM-neutrino couplings, we need to consider more interaction possibilities with realistic neutrino mass orderings.


In this paper, we shall study the cosmological constraints on the temperature-dependent DM-neutrino interactions. We begin our discussion by introducing a specific dimension-six operator, which describes the interaction between DM particles and three active left-handed neutrinos. It turns out that this coupling would give rise to the DM-neutrino scattering cross section which, in the high-temperature limit, can be approximated to be proportional to the neutrino temperature squared, as considered in Ref.~\cite{Wilkinson:2014ksa}. However, as the universe cools down, neutrinos become non-relativistic, leading to a more complicated temperature dependence and different dynamics. We will investigate this mass effect in our temperature-dependent setup, which should be more pronounced than the temperature-independent case in Ref.~\cite{Mosbech:2020ahp}. To this aim, we will follow Ref.~\cite{Mosbech:2020ahp} to derive the full Boltzmann hierarchy for massive neutrinos in the presence of temperature-dependent interactions with DM. Especially, we shall present the detailed calculation of the collision terms in either the neutrino Boltzmann hierarchy or the DM Euler equation. Then we implement these modified Boltzmann equations into the general solver \texttt{CLASS}~\cite{Blas:2011rf}, and obtain the constraints on our proposed DM-neutrino interaction from the latest cosmological data with the Monte Carlo Markov Chain (MCMC) sampler \texttt{Cobaya}~\cite{Torrado:2020dgo}.  Note that when the neutrino modes enter the horizon deeply, it is argued~\cite{Lesgourgues:2011rh} that the full neutrino hierarchy can be simplified by using the fluid approximation. In Ref.~\cite{Mosbech:2020ahp}, the neutrino fluid equations were obtained by fitting the integrated rate as a function of the massive neutrino equation of state from the simulated data. In the present work, we shall derive the neutrino fluid equations by firstly averaging the perturbation moments to yield the macroscopic quantities like the density contrast, velocity divergence, and shear stress, so that the transport coefficients could be defined and calculated directly by \texttt{CLASS} without any use of numerical simulations. What is more, the three neutrino masses in our benchmark model obey the normal ordering, {\it i.e.}, two massless flavors and one massive one, which is more favored than the degenerate mass model studied in Ref.~\cite{Mosbech:2020ahp}. Nevertheless, we shall still perform the fits in the degenerate mass case for comparison. Finally, it was noticed in Refs.~\cite{Brax:2023tvn,Giare:2023qqn,Zu:2025lrk} that, when including the latest ACT-DR6 data~\cite{ACT:2025fju,ACT:2025tim} and using the log-flat prior for the DM-neutrino scattering parameter, the fits exhibited a slight preference to the non-zero DM-neutrino interaction. Thus, we will incorporate the ACT-DR6 data in our parameter exploration and study the effects from different prior choices.  


This paper is structured as follows. Sec.~\ref{TH} presents our model for the DM-$\nu$ interaction based on a dimension-six effective operator. After reviewing the derivation of Boltzmann equations in the $\Lambda$CDM model, we detail the calculation of the Boltzmann hierarchy for neutrinos in our model, paying attention to the computation of the collision terms from the microscopic DM-neutrino coupling. 
Sec.~\ref{Meth} is devoted to discussing the influence of DM-neutrino scatterings on the CMB and the matter power spectra. In Sec.~\ref{NR}, after presenting the analysis methods, numerical codes, and datasets, we give our numerical results for various model setups. 
Finally, Sec.~\ref{conclusion} summarizes the main conclusions. Detailed derivations of the modified Boltzmann equations and the fluid approximation 
are given in Appendix~\ref{Derivation} and Appendix~\ref{FA}, respectively. We also present the triangular plot for all the setups considered in this work in Appendix~\ref{Tri}.

\section{A Model for DM-Neutrino Interactions and Modified Boltzmann Equations}\label{TH}
Temperature-dependent interactions between neutrinos and DM have been discussed in Refs.~\cite{Wilkinson:2014ksa,Brax:2023tvn}, where the quadratic temperature dependence of the interaction cross section has been considered. As shown later, it can be achieved by a dimension-six operator describing DM-neutrino scatterings only for massless neutrinos. Nevertheless, neutrino oscillation experiments tell us that neutrinos possess nonzero masses, which make the interaction strength as a complicated function of temperature, especially at low temperatures. Unfortunately, the exact behavior of the scattering cross section is not universal and depends on the concrete microscopic model details. In this section, we shall first specify a dimension-six DM-neutrino coupling operator under the philosophy of effective field theory (EFT). Then we shall derive the temperature reliance of the interaction cross section, which will be further used to yield the associated modified Boltzmann equations for DM and neutrinos.

\subsection{A Effective Coupling and DM-Neutrino Interaction Cross Section}\label{SecLag}
Let us begin with the following effective Lagrangian describing the scatterings of the DM particle $\chi$ and active left-handed neutrinos $\nu_L$   
\begin{align}\label{lagrangian}
	\mathcal{L}_{\chi\nu} = \frac{1}{\Lambda ^2}\bar{\chi}\gamma^{\mu}\chi\bar{\nu_{L}}\gamma_{\mu}\nu_{L} ,
\end{align}
where $\Lambda$ is the cutoff scale and DM is a Dirac fermion $\chi$. 
Note that renormalizable operators with mass dimensions less than four can only give rise to a constant cross section for DM-$\nu$ scatterings at low temperatures. This is the reason why we choose the dimension-six operator in Eq.~\eqref{lagrangian}, which is the lowest order to produce the temperature effects. From the EFT perspective, such operators should provide the largest temperature-related contribution.
  

Given the DM-neutrino interactions in Eq.~\eqref{lagrangian}, we can easily yield the following spin-averaged amplitude squared in the center-of-mass frame: 
\begin{align}\label{scattering_amplitude}
	|\mathcal{M}|^2 \approx \frac{4m_{\chi}^2}{\Lambda^4} \left(2E_{\nu}^2- E^\prime_{\nu}E_\nu-\mathbf{p}\cdot\mathbf{p'} \right),
\end{align}
where $E_\nu$ and $E^\prime_{\nu}$ represent energies of the incoming and outgoing neutrinos, respectively. $m_\chi$ is the DM mass, and $\mathbf{p}$ and $\mathbf{p}^\prime$ denote corresponding neutrino momenta. 
The scattering cross-section can be given by:
\begin{align}\label{scattering_cross_section}
	\sigma_{\chi-\nu} \approx \frac{E_\nu^2}{4\pi \Lambda^4},
\end{align}
By defining today's scattering cross section as $\sigma_{\chi-\nu}^0$, we can rewrite the above formula as
\begin{align}\label{scattering_cross_section_a}
	\sigma_{\chi-\nu} = \sigma_{\chi-\nu}^0 \frac{E_\nu^2 }{E_{\nu,0}^2},
\end{align}
where $E_{\nu,0}$ is today's neutrino energy after the cosmological redshift. If neutrinos are treated as massless, the ratio ${E_\nu^2 }/{E_{\nu,0}^2}$ becomes $a^{-2}$ as predicted by cosmological expansion, where $a$ is the scale factor in the Friedmann-Robertson-Walker (FRW) metric. However, the existence of neutrino masses would lead to significant deviations from this behavior, especially when massive neutrinos become non-relativistic at low temperatures. 

By taking Eq.\eqref{scattering_cross_section_a} into Eq.~\eqref{scattering_amplitude} , we can express the DM-$\nu$ scattering amplitude squared as
\begin{align}\label{scattering_amplitude_a}
	|\mathcal{M}|^2 \approx16\pi m_\chi^2 \sigma_{\chi-\nu}^0\frac{E_\nu^2 }{E_{\nu,0}^2} \left(2-\frac{E_\nu'}{E_\nu}+\frac{\mathbf{p}\cdot\mathbf{p'}}{E_\nu^2}\right)\,,
\end{align}
which would be used to derive the  corresponding collision term in the Boltzmann equations in Sec.~\ref{SecBoltzmann}.

\subsection{The Boltzmann Equations in the Standard Cosmology}\label{MBE}
Before going into the calculation in our DM-neutrino interaction model, we firstly review the derivation of the Boltzmann equations for DM and neutrinos in the $\Lambda$CDM model, 
in order to set up our notations and conventions.


We start with the general relativistic Boltzmann equations~\cite{Ma:1995ey,Oldengott:2017fhy}, describing the evolution of phase-space distribution functions $f$ for a given species
\begin{align}\label{boltzmann}
P^{\alpha} \frac{\partial f}{\partial x^{\alpha}} - \Gamma_{\alpha\beta}^{\gamma} P^{\alpha} P^{\beta} \frac{\partial f}{\partial P^{\gamma}} = m \left( \frac{\partial f}{\partial \tau} \right)_{C},
\end{align}
where $P^{\alpha}$ and $m$ represents the particle physical four-momentum and mass, while $\Gamma_{\alpha\beta}^{\gamma}$ and $\tau$ are Christoffel symbols and conformal time for a given cosmological background geometry. 
The right-hand side of this equation denotes the collision term. 
In order to implement the perturbation theory, it is convenient to decompose the phase-space distribution into the background contribution $ f^{(0)}(p)$ and its perturbation $\Psi(\mathbf{x},\mathbf{P}, \tau)$
\begin{align}\label{perturbation}
f(\mathbf{x}, \mathbf{P}, \tau) = f^{(0)}\left(p\right) [1+\Psi(\mathbf{x}, \mathbf{P}, \tau)],
\end{align}
where $p = |\mathbf{P}|$ is the magnitude of momentum. 
Since we are interested in massive neutrinos, 
their background should obey the Fermi-Dirac distribution:
\begin{align}\label{fermi_dirac}
f_s^{(0)}\left(p\right) = g_s\frac{1}{e^{E/T_s} + 1},
\end{align}
in which the subscript $s$ denotes the fermion species, while $g_s$ and $T_s$ are the corresponding spin degrees of freedom and temperature. In the FRW background metric and the conformal Newtonian gauge for metric perturbations, the Boltzmann equation for $\Psi$ reads:
\begin{align}\label{boltzmann_newtonian}
\frac{\partial \Psi}{\partial \tau} + i \frac{p}{E} \left( \mathbf{k} \cdot \hat{\mathbf{n}} \right) \Psi + \frac{d \ln f_s^{(0)}\left(p\right)}{d \ln p} \left[ \dot{\phi} - i \frac{E}{p} \left( \hat{\mathbf{k}} \cdot \hat{\mathbf{n}} \right) \psi \right] = \frac{1}{f_s^{(0)}} \left( \frac{\partial f}{\partial \tau} \right)_C\,, 
\end{align}
where $\phi$ and $\psi$ are the conformal Newtonian potentials, $\hat{\mathbf{n}}$ is the direction of the fermion momentum, and $\mathbf{k}$ is the Fourier transform variable for coordinates $\mathbf{x}$, respectively. Here $\mathbf{P}$ and $p$ denote the particle physical momentum and its amplitude, so that the physical energy is given by $E=\sqrt{p^2+m^2}$. 

In the $\Lambda$CDM model, DM is treated as a pressureless perfect fluid and interacts with other particles only through gravity, so that 
its Boltzmann equations are simply given by
\begin{align}\label{dark_matter_fluid}
\dot{\delta}_{\chi} &= -\theta_{\chi} + 3\dot{\phi}, \\
\dot{\theta}_{\chi} &= -\frac{\dot{a}}{a} \theta_{\chi} + k^2 \psi,
\end{align}
which are derived by performing a Legendre decomposition on Eq.~\eqref{boltzmann_newtonian} and integrating over the momentum $p$. Here the DM density contrast $\delta_{\chi}$ and its velocity divergence $\theta_{\chi}$ evolve as functions of $\tau$. Note that these equations can be derived by assuming that the nonrelativistic DM particles follow the distribution function
\begin{eqnarray}
 g(\mathbf{q}) \simeq n_\chi (2\pi)^3 \delta^3(\mathbf{q}-m_\chi \mathbf{v}_\chi)\,,
\end{eqnarray} 
in which $n_\chi$ denotes the DM number density and $\mathbf{v}_\chi(\mathbf{x})$ is its three-dimensional velocity field. With the further assumption that the velocity $\mathbf{v}_\chi$ is irrotational, {\it i.e.}, $\nabla \times \mathbf{v}_\chi = 0$, its direction should be parallel to that of the coordinate Fourier transform $\mathbf{k}$. As a result, the only nontrivial information about the DM velocity is captured by the its divergence $\nabla \cdot \mathbf{v}_\chi$. After the Fourier transform, this quantity can be defined as $\theta_\chi \equiv ikv_\chi$. 

On the other hand, since heavy neutrinos become non-relativistic as the temperature drops down, we should take into account their mass effect and solve the dynamics in a momentum-dependent way. By expanding the neutrino perturbation $\Psi(\mathbf{x},\mathbf{P},\tau)$ in terms of the spherical harmonics $P_l(\mu)$ with $\mu\equiv \hat{\mathbf{P}}\cdot \hat{\mathbf{k}}  =\hat{\mathbf{P}}\cdot\hat{\mathbf{v}}_\chi$ where $\hat{\mathbf{P}}$ and  $\hat{\mathbf{v}}_\chi$ represent the directions of neutrino momenta and DM fluid velocities, the full Boltzmann hierarchy for collisionless massive neutrinos in the conformal Newtonian gauge is given by~\cite{Ma:1995ey}
\begin{align}
\frac{\partial \Psi_0}{\partial \tau} &= -\frac{pk}{E_\nu\left(p\right)} \Psi_1 - \dot{\phi} \frac{d \ln f^{(0)}\left(p\right)}{d \ln p}\,, \label{Bolt0}\\
\frac{\partial \Psi_1}{\partial \tau} &= \frac{1}{3} \frac{pk}{E_\nu\left(p\right)} \left( \Psi_0 - 2\Psi_2 \right) - \frac{E_\nu\left(p\right) k}{3p} \psi \frac{d \ln f^{(0)}\left(p\right)}{d \ln p}\,,  \label{Bolt1}\\
\frac{\partial \Psi_l}{\partial \tau} &= \frac{1}{2l + 1} \frac{pk}{E_\nu\left(p\right)} \left( l \Psi_{l-1} - (l+1) \Psi_{l+1} \right), \quad l \geq 2\,, \label{BoltL}
\end{align}
where $\Psi_l$ denote the $l$-th Legendre moments of the neutrino phase-space perturbation $\Psi$.

\subsection{The Modified Boltzmann Equations for Temperature-Dependent DM-Neutrino Couplings}\label{SecBoltzmann}
The discussion in the previous subsection is only applicable for the case with collisionless DM and neutrinos, which is one of basic assumptions in the standard cosmological model. However, when the DM-neutrino interaction represented by Eq.~\eqref{lagrangian} is added to the theory, the DM and neutrino Boltzmann equations should be modified by collision terms. Let us begin with the evaluation of the following collision integral for massive neutrinos with momentum $p$ 
\begin{align}\label{collision}
C\left(p\right) &= \frac{1}{2E_\nu\left(\mathbf{p}\right)} \int \frac{d^3\mathbf{p'}}{(2\pi)^3 2E_\nu\left(\mathbf{p'}\right)} \frac{d^3\mathbf{q}}{(2\pi)^3 2E_\chi(\mathbf{q})} \frac{d^3\mathbf{q'}}{(2\pi)^3 2E_\chi(\mathbf{q'})} \\
&\quad \times (2\pi)^4 |\mathcal{M}|^2 \delta^4 \left( q + p - q' - p' \right) \left[ g(\mathbf{q}') f(\mathbf{p}') (1 - f(\mathbf{p})) - g(\mathbf{q}) f(\mathbf{p}) (1 - f(\mathbf{p}')) \right].\nonumber 
\end{align}
where $f(\mathbf{p})$ and $g(\mathbf{q})$ denote the phase-space distributions for neutrinos and DM particles, respectively. $\mathbf{q}$ and $\mathbf{q}^\prime$ are the incoming and outgoing momenta for DM, while $\mathbf{p}$ and $\mathbf{p}^\prime$ denote the corresponding neutrino momenta. As shown in Sec.~\ref{SubCT}, substituting the DM-$\nu$ squared amplitude in Eq.~\eqref{scattering_amplitude_a} into the above expression gives 
\begin{align}\label{eq:collision_integral}
C\left(p\right) &=  a^{-1}C_\chi \left(p\right)
\Big[f_0^{(1)}\left(p\right) + \frac{p^2}{E^2_\nu\left(p\right)}P_1(\mu)f_1^{(1)}\left(p\right)-f^{(1)}(\mu,p) \nonumber\\
&-v_\chi \mu E_\nu\left(p\right)\left(1-\frac{p^2}{3E^2_\nu\left(p\right)}\right)\frac{df^{(0)}}{dp}\Big]\,,
\end{align}
where we have adopted the same notation as in Ref.~\cite{Mosbech:2020ahp} and similarly defined
\begin{align}\label{C_nudm}
    C_\chi \left(p\right) = \frac{a \sigma_{\chi-\nu}^0 n_\chi pE_{\nu}\left(p\right)}{E^2_{\nu,0}\left(p\right)}\,.
\end{align}
As argued below Eq.~\eqref{scattering_cross_section_a}, the factor ${p E_{\nu}\left(p\right)}/{E^2_{\nu,0}\left(p\right)}$ implies that the collision term $C(p)$ varies as the neutrino temperature via the cosmological redshift. 

It is convenient to re-parametrize the DM-massive neutrino cross section with  the following dimensionless quantity 
\begin{align}
    u_{\chi-\nu}^0=\frac{\sigma_{\chi-\nu}^0}{\sigma_\mathrm{Th}}\left(\frac{m_\chi}{100~\mathrm{GeV}}\right)^{-1},
\end{align}
where $\sigma_\mathrm{Th}$ denotes the cross section of the conventional Thomson scatterings. 
As a result, the collision coefficient $C_\chi(p)$ of Eq.~\eqref{C_nudm} can be rewritten as
\begin{align}\label{C_nudm2}
     C_\chi \left(p\right) = a u_{\chi-\nu}^0\frac{\sigma_\mathrm{Th}\rho_\chi}{100~\mathrm{GeV}}\frac{p E_{\nu}\left(p\right)}{E_{\nu,0}^2\left(p\right)}.
\end{align}
In order to explore the impact of collision term in Eq.~\eqref{eq:collision_integral} on the neutrino Boltzmann hierarchy, we need to expand it in terms of the spherical harmonics. As derived in Sec.~\ref{SubBH}, such a collision term can only modify the associated Boltzmann equations for moments $l\geqslant 1$, with the explicit results given by   
\begin{align}
\frac{\partial \Psi_0}{\partial \tau} &= -\frac{pk}{E_\nu\left(p\right)} \Psi_1 - \dot{\phi} \frac{d \ln f^{(0)}\left(p\right)}{d \ln p}\,, \label{BoltNu0} \\
\frac{\partial \Psi_1}{\partial \tau} &= \frac{1}{3} \frac{pk}{E_\nu\left(p\right)} \left( \Psi_0 - 2\Psi_2 \right) - \frac{E_\nu\left(p\right) k}{3p} \psi \frac{d \ln f^{(0)}\left(p\right)}{d \ln p} \nonumber \\
&\quad -C_\chi\left(p\right)\left(1-\frac{p^2}{3E^2_\nu\left(p\right)}\right)\left(\Psi_1+\frac{1}{3}iv_\chi E_\nu\left(p\right)\frac{1}{f^{(0)}\left(p\right)}\frac{df^{(0)}}{dp}\right)\,, \label{BoltNu1} \\
\frac{\partial \Psi_l}{\partial \tau} &= \frac{1}{2l + 1} \frac{pk}{E_\nu\left(p\right)} \left[ l \Psi_{l-1} - (l+1) \Psi_{l+1}\right] - C_\chi\left(p\right)\Psi_l, \quad l \geq 2\,. \label{BoltNu2}
\end{align}
It is interesting to note that the first equation with $l=0$ remains unmodified, as the presence of DM-neutrino interactions does not induce the energy transfer at the linear order. 

The DM-neutrino interaction also affects the Boltzmann equations for DM, which, in contrast to that for massive neutrinos, are usually represented by the continuity and Euler equations for the DM fluid. Rather than directly deriving the modifications for these DM equations, we shall take another routine to firstly integrate the hierarchy of massive neutrinos over their momentum space so as to obtain the corrected neutrino continuity and Euler equations, and then alter the DM Euler equations according to the momentum conservation in the neutrino-DM coupled fluid. Let us begin by recalling the relationships of neutrino perturbations $\Psi_0$, $\Psi_1$ and $\Psi_2$ with the density contrast $\delta_\nu$, the velocity divergence $\theta_\nu$ and the shear stress $\sigma_\nu$~\cite{Ma:1995ey}:
\begin{align}
\rho_\nu \delta_\nu &= 4\pi \int p^2 dp E_\nu(p) f^{(0)}(p)\Psi_0\,,\label{DefRho}\\
\left(\rho_\nu+P_\nu\right)\theta_\nu &= 4\pi k \int p^2 dp \, pf^{(0)}(p)\Psi_1\,, \label{DefTheta}\\
\left(\rho_\nu+P_\nu\right)\sigma_\nu &= \frac{8\pi}{3} \int p^2 dp \, \frac{p^2}{E_\nu(p)}f^{(0)}(p)\Psi_2\,, \label{DefSigma}
\end{align}
where $\rho_\nu$ and $P_\nu$ are the averaged energy density and pressure of the massive neutrino fluid, with their definitions given by 
\begin{align}\label{energy_density_pressure}
\rho_\nu &= 4\pi\int p^2 dp \, E_\nu(p) f^{(0)}(p), \\
P_\nu &= \frac{4\pi}{3} \int p^2 dp \, \frac{p^2}{E_\nu(p)} f^{(0)}(p).
\end{align}
Based on these definitions, the integration of the first three multipoles of the neutrino Boltzmann hierarchy in Eqs.~\eqref{BoltNu0}, \eqref{BoltNu1}, and \eqref{BoltNu2} over the momentum $p$ can give us equations for the neutrino density contrast $\delta_\nu$, velocity divergence $\theta_\nu$ and shear $\sigma_\nu$. As shown in Eq.~\eqref{BoltNu0}, the interaction only introduces DM-$\nu$ scatterings which do not make any change for the equation of $\delta_\nu$. Thus, here we only present the following modified equations for $\theta_\nu$ and $\sigma_\nu$
\begin{eqnarray}
\dot{\theta}_{\nu} &=& [\dots] - k \frac{ \int p^2 dp \, p f^{(0)}(p)  C_\chi \left(p \right)   \left( 1 - \frac{p^2}{3E^2_\nu(p)} \right)
  \left( \Psi_1 + \frac{\theta_\chi}{3k} E_\nu(p) \frac{1}{f^{(0)}(p)} \frac{df^{(0)}}{dp} \right)}{
    \int p^2 dp \, E_\nu(p) f^{(0)}(p) 
    + \frac{1}{3} \int p^2 dp \, \frac{p^2}{E_\nu(p)} f^{(0)}(p)} \,, \label{NuTheta0} \\
 \dot{\sigma_\nu} &=& [\dots] - \frac{2}{3}\frac{\int p^2 dp \, \frac{p^2}{E_\nu(p)}f^{(0)}(p)C_\chi \left(p \right)\Psi_2}
{\int p^2 dp \, E_\nu(p) f^{(0)}(p) 
+ \frac{1}{3} \int p^2 dp \, \frac{p^2}{E_\nu(p)} f^{(0)}(p)} \,. \label{NuSigma0}
\end{eqnarray}
where the terms in the brackets $[\dots]$ represent those which keep the same as the non-interacting case. 

By virtue of momentum conservation in the DM-massive neutrino fluid, we can derive the modified Euler equation for the DM sector as:
\begin{align}\label{theta_chi_dot}
\dot{\theta}_{\chi} = [\dots] + K_\chi k 
\frac{
    \int p^2 dp \, p f^{(0)}(p) \,
    C_\chi \left(p \right)
    \left( 1 - \frac{p^2}{3E^2_\nu(p)}\right)
    \left( \Psi_1 + \frac{\theta_\chi}{3k} E_\nu(p) \frac{1}{f^{(0)}(p)} \frac{df^{(0)}}{dp} \right)
}{
    \int p^2 dp \, E_\nu(p) f^{(0)}(p) 
    + \frac{1}{3} \int p^2 dp \, \frac{p^2}{E_\nu(p)} f^{(0)}(p)
}\,,
\end{align}
in which the factor $K_\chi$ coming from the momentum conservation is
\begin{align}\label{K_chi}
    K_{\chi} \equiv \frac{\rho_{\nu} + P_{\nu}}{\rho_{\chi}} = \frac{(1 + w_{\nu})\rho_{\nu}}{\rho_{\chi}}\,.
\end{align}
We have also defined $w_{\nu} = P_{\nu}/\rho_{\nu}$ as the equation of state parameter for the neutrino fluid, which is not constant for massive neutrinos and varies with the temperature. $\rho_{\chi}\equiv m_\chi n_\chi$ is the DM energy density. For a more detailed derivation of the modified Boltzmann equations for massive neutrinos and DM, interested readers are invited to refer to Appendix~\ref{Derivation}.

Note that when the mode wavelengths become deeply inside the Hubble radius, only the first three multipoles are sourced by the gravitational potential, while higher multipoles experience damped oscillations, leading to effective decoupling between the lowest three and higher multipole modes. In this regime, we can apply the fluid approximation~\cite{Mosbech:2020ahp,Lesgourgues:2011rh,Blas:2011rf} to simplify the numerical calculation of the Boltzmann hierarchy. Under this approximation, we only need to take into account the first three multipoles. In particular, by firstly averaging $\Psi_1$ and $\Psi_2$ over momentum $p$ to obtain $\theta_\nu$ and $\sigma_\nu$,  Eqs.~\eqref{NuTheta0}, \eqref{NuSigma0} and \eqref{theta_chi_dot} can be transformed into the following fluid equations
\begin{align}
	\dot{\theta}_{\nu} = [\dots] &- \frac{
		\int p^2 dp \, f^{(0)}(p)\, 
		C_\chi\left(p\right)
		\left( 1 - \frac{p^2}{3E^2_\nu(p)}\right)
	}{\int p^2 dp \, f^{(0)}(p)}\theta_\nu\nonumber\\
	\quad &-\frac{1}{3}\frac{
		\int p^2 dp \, p f^{(0)}(p)\, 
		C_\chi\left(p\right)
		\left( 1 - \frac{p^2}{3E^2_\nu(p)} \right)
		\frac{E_\nu(p)}{f^{(0)}(p)} \frac{df^{(0)}}{dp}
	}{
		\int p^2 dp \, E_\nu(p) f^{(0)}(p) 
		+ \frac{1}{3} \int p^2 dp \, \frac{p^2}{E_\nu(p)} f^{(0)}(p)
	}\theta_\chi\,, \label{thetaNuF} \\
	\dot{\sigma}_\nu = [\dots] &- \frac{\int p^2 dp \, f^{(0)}(p)C_\chi \left(p \right)}
	{\int p^2 dp \,  f^{(0)}(p) 
	}\sigma_\nu\,, \label{sigmaNuF} \\
	\dot{\theta}_{\chi} = [\dots] &+K_\chi k \frac{
		\int p^2 dp \, f^{(0)}(p)\, 
		C_\chi\left(p\right)
		\left( 1 - \frac{p^2}{3E^2_\nu(p)}\right)
	}{\int p^2 dp \, f^{(0)}(p)}\theta_\nu\nonumber\\
	\quad &+\frac{1}{3}K_\chi k \frac{
		\int p^2 dp \, p f^{(0)}(p)\, 
		C_\chi\left(p\right)
		\left( 1 - \frac{p^2}{3E^2_\nu(p)} \right)
		\frac{E_\nu(p)}{f^{(0)}(p)} \frac{df^{(0)}}{dp}
	}{\int p^2 dp \, E_\nu(p) f^{(0)}(p) 
		+ \frac{1}{3} \int p^2 dp \, \frac{p^2}{E_\nu(p)} f^{(0)}(p)
	}\theta_\chi\,, \label{thetaChiF}
\end{align}
while the fluid equations for $\delta_\nu$ and $\delta_\chi$ are kept intact as the non-interacting case. 
Note that the fluid equations obtained here are different from those in Ref.~\cite{Mosbech:2020ahp}. The details of the derivation are elaborated in Appendix~\ref{FA}. 




\section{Numerical Analysis}\label{Meth}
The cosmological Boltzmann hierarchies with the temperature-dependent DM-$\nu$ interactions are numerically solved with a modified version of the \texttt{CLASS} code~\cite{Lesgourgues:2011rh,Blas:2011rf}, where we have incorporated DM-neutrino interactions for both the massless and massive neutrinos based on the public code in Refs.~\cite{Stadler:2019dii,Mosbech:2020ahp}\footnote{A publicly available version can be found at \url{https://github.com/MarkMos/CLASS_nu-DM}; for more details, see Refs.~\cite{Stadler:2019dii,Mosbech:2020ahp}.}. We have also implemented in the code the fluid equations according to Eqs.~\eqref{thetaNuF}, \eqref{sigmaNuF} and \eqref{thetaChiF} when the fluid approximation is applicable. Furthermore, the existence of neutrino oscillations has revealed that active neutrinos should have  tiny but nonzero masses~\cite{ParticleDataGroup:2024cfk,Super-Kamiokande:1998kpq,MACRO:2001fie,Soudan2:2003qqa,IceCube:2019dqi,ANTARES:2018rtf,K2K:2006yov,MINOS:2014rjg,MINOS:2020llm,OPERA:2018nar,Rubbia:2011ft,T2K:2011ypd,NOvA:2017abs,KamLAND:2002uet,DoubleChooz:2011ymz,DayaBay:2012fng,RENO:2012mkc,JUNO:2021vlw}. Thus, we shall work in the normal ordering as our benchmark scenario by including two massless and one massive neutrino flavors, which is a standard setup commonly adopted in recent cosmological studies~\cite{ACT:2025fju,Planck:2019nip,DESI:2025zgx}. 

\begin{figure}[htbp]
	\centering
	\includegraphics[width=0.8\textwidth]{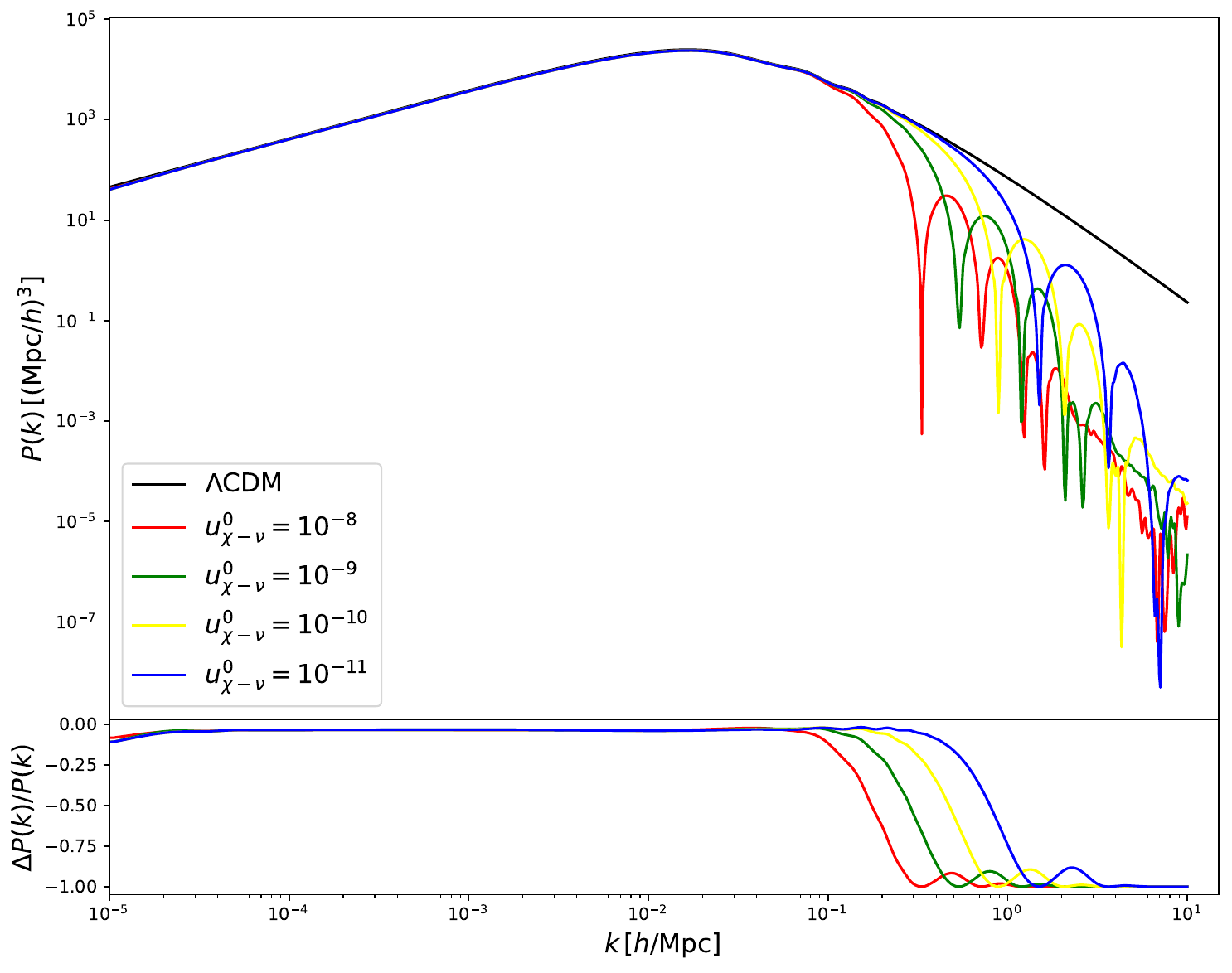}
	\caption{The upper panel shows the matter power spectra for different values of the interaction parameter $u^0_{\chi-\nu}$, with the cosmological parameters fixed by the Planck best-fit values~\cite{Planck:2018vyg}. 
	The bottom panel shows the relative differences, compared to the $\Lambda$CDM model.}
	\label{fig:Pk}
\end{figure}

Fig.~\ref{fig:Pk} shows the effect of DM-neutrino interactions on the matter power spectrum. When the DM-neutrino scatterings become efficient, the spectrum displays damped oscillations at small scales~\cite{Wilkinson:2014ksa,Oldengott:2014qra,Mosbech:2020ahp,Stadler:2019dii}., which arises from the dynamic competition between the DM gravitational clustering and the neutrino-DM fluid pressure. Moreover, such the DAO in the DM-$\nu$ fluid would suppress the DM fluctuation amplitude for modes with large wavenumbers $k$ compared to the non-interacting case. 

\begin{figure}[htbp]
	\centering
	\includegraphics[width=0.6\textwidth]{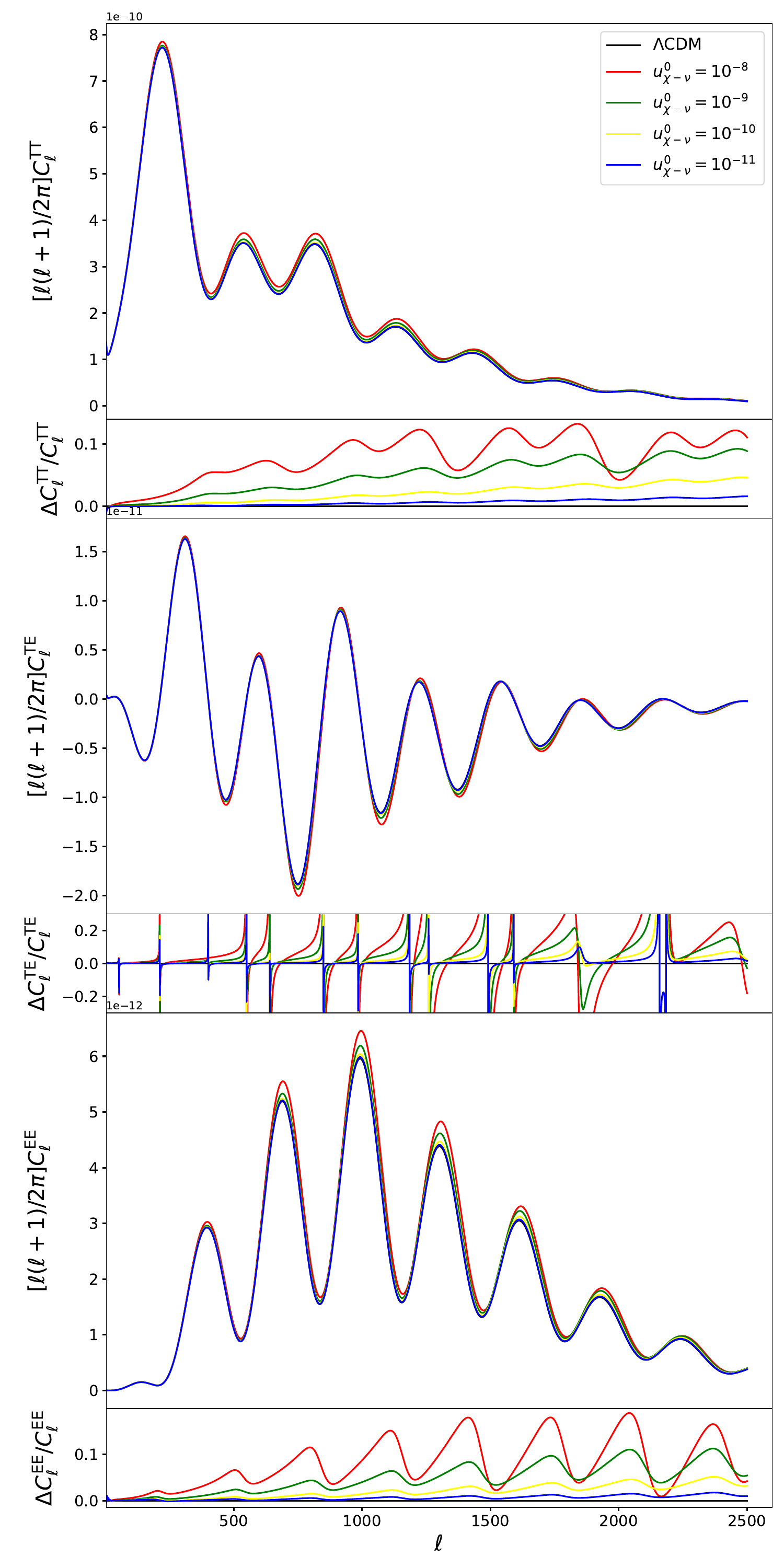}
	\caption{The first, third, and fifth panels show the CMB TT, TE, and EE spectra for different values of $u^0_{\chi-\nu}$, calculated using the cosmological parameters from Planck TTTEEE best-fit data~\cite{Planck:2018vyg}, 
	while the second, fourth, and sixth panels represent the associated relative differences, compared to the $\Lambda$CDM model.}
	\label{fig:CMB}
\end{figure}

The impacts of the DM-$\nu$ scattering on the CMB TT, TE and EE spectra are illustrated in Fig.~\ref{fig:CMB}. Notably, the anisotropic CMB signals primarily originate from the evolution of the coupled baryon-photon fluid in the early universe, the properties of which are also affected by the neutrino free-streaming and the DM clustering. Note that the cosmological perturbations can be decomposed into slow and fast modes~\cite{Voruz:2013vqa,Weinberg2008cosmology}. In the absence of DM-neutrino interactions, the perturbations in the baryon-photon fluid and those for neutrinos can be viewed as fast modes, while DM are dominated by slow ones. In this case, only neutrinos exert a significant influence on the baryon-photon fluid via the gravitational attraction. However, the presence of the DM-neutrino coupling causes DM to exhibit rapid DAOs instead of slow clustering, so that DM can also contribute to the fast modes. In this way, the peak amplitudes of the CMB spectra are enhanced through the gravitational boosting effect. Additionally, when DM and neutrinos are tightly coupled, their sound speed is lower than that of the baryon-photon fluid, leading to the drag effect. This results in a slight reduction in the acoustic scale, causing the peak positions to shift toward higher multipoles~\cite{Wilkinson:2014ksa,Mangano:2006mp,Mosbech:2020ahp}, which can be clearly seen in the TT, TE, and EE spectra in Fig.~\ref{fig:CMB}. 


\section{Fitting Results}\label{NR}
To precisely measure the impact of the DM-neutrino interactions on cosmological parameter measurements, we shall employ the Markov-Chain Monte-Carlo (MCMC) sampling technique to explore the posterior distributions over the parameter spaces. As mentioned before, we are working in the scenario with the normal ordering of neutrino masses, namely two masssless and one massive flavors. In our benchmark model, the total neutrino mass is fixed to be $\sum m_\nu = 0.06$~eV. Since we have taken into account the interactions between DM and all neutrino flavors, the benchmark parameter space is enlarged by including the interaction parameter $u^0_{\chi-\nu}$ together with the conventional six $\Lambda$CDM ones: the amplitude of primordial scalar perturbations $\log(10^{10} A_s)$, the scalar spectral index $n_s$, the angular size of the horizon at the last scattering surface $100 \theta_s$, the baryon energy density $\omega_b=\Omega_b h^2$, the cold dark matter energy density $\omega_c=\Omega_c h^2$, and the optical depth to reionization $\tau_{\mathrm{reio}}$. The linear flat prior for $u^0_{\chi-\nu}$ is adopted in our benchmark fitting.


Besides the above benchmark setup, we further consider the effects of some variations of model assumptions and fitting procedures. Firstly, the DM-$\nu$ interaction model with a constant cross section was investigated in Ref.~\cite{Mosbech:2020ahp} where all neutrinos were assumed to be massive and degenerate with $\sum m_\nu = 0.06$~eV. In order for comparison, we will study such a degenerate neutrino mass scenario but with our temperature-dependent interactions. Furthermore, to comprehensively study the neutrino mass effects on our model with normal ordering, we shall vary the neutrino mass $\sum m_\nu$ and treat it as a free parameter in a separate fitting. Finally, Ref.~\cite{Diacoumis:2018ezi} pointed out that the one-sided constraint on the DM-neutrino interaction $u^0_{\chi-\nu}$ derived from the CMB data depended on the prior distribution of this parameter. Especially, if the alternative logarithmic flat prior for $u^0_{\chi-\nu}$ was adopted, the fits in Ref.~\cite{Diacoumis:2018ezi} showed slight preference to the non-zero DM-$\nu$ scattering cross section. Therefore, we shall also consider a scenario where a logarithmic flat prior of $u^0_{\chi-\nu}$ is taken while other parameter priors are kept as the benchmark model. The purpose of the last study is to quantify the impact of prior choices of $u^0_{\chi-\nu}$ on the numerical analysis. In sum, the prior distributions for all sampled parameters involved in our analysis are listed in Table~\ref{tab:uniform_priors}. Note that the effective number of neutrino species is always fixed to be $N_\mathrm{eff}=3.044$ according to recent studies~\cite{Akita:2020szl,Froustey:2020mcq,Bennett:2020zkv}. 
\begin{table}[htb!]
	\centering
	\caption{List of model parameter priors for different setups. The columns labeled by 'linear' and 'logarithmic' denote the setups with the linear flat and logarithmic flat priors for the interaction parameter $u^0_{\chi-\nu}$ ($\log_{10} u^0_{\chi-\nu}$), respectively.  The column labeled by 'varying mass' represents the case in which the total neutrino mass is also sampled. Note that the linear setup is used for our benchmark scenario with the normal neutrino mass ordering and degenerate masses. }
	\begin{tabular}{lccc}
		\hline
		Parameter & linear \,\, & varying mass \,\, & logarithmic \\
		\hline
		$u^0_{\chi-\nu}(\log_{10}u^0_{\chi-\nu})$ & $[0,\: 10^{-10}]$ & $[0,\: 10^{-10}]$ & $[-18,\: -10]$ \\
		$\log(10^{10} A_s)$ & $[1.61,\: 3.91]$ & $[1.61,\: 3.91]$ & $[1.61,\: 3.91]$ \\
		$n_s$ & $[0.8,\: 1.2]$ & $[0.8,\: 1.2]$ & $[0.8,\: 1.2]$ \\
		$100\theta_s$ & $[0.5,\: 10]$ & $[0.5,\: 10]$ & $[0.5,\: 10]$ \\
		$\Omega_b h^2$ & $[0.005,\: 0.1]$ & $[0.005,\: 0.1]$ & $[0.005,\: 0.1]$ \\
		$\Omega_{c} h^2$ & $[0.005,\: 0.1]$ & $[0.005,\: 0.1]$ & $[0.005,\: 0.1]$ \\
		$\tau_\mathrm{reio}$ & $[0.01,\: 0.8]$ & $[0.01,\: 0.8]$ & $[0.01,\: 0.8]$ \\
		$\sum m_\nu$ [eV] & $0.06$ & $[0.0,\: 0.99]$ & $0.06$\\
		\hline
	\end{tabular}
	\label{tab:uniform_priors}
\end{table}

We have performed our MCMC samplings for all DM-$\nu$ interaction scenarios with the widely-used code \texttt{Cobaya}~\cite{Torrado:2020dgo}. 
Then the obtained Monte Carlo samples are analyzed with the Python package \texttt{GetDist}~\cite{Lewis:2019xzd} which enables us to yield the statistical summaries and visualize the posterior distributions. The convergence of the chains is assessed by using the Gelman-Rubin criterion~\cite{Gelman:1992zz} with the convergence threshold chosen to be $R-1\lesssim 0.01$. In our numerical analysis, the following baseline datasets are used:

\begin{itemize}
    \item The Planck 2020 (PR4) high-$\ell$ TT, TE, and EE likelihoods are combined with Planck 2018 low-$\ell$ EE and TT. Planck PR4 corresponds to the $\texttt{CamSpec}$ likelihoods based on the Planck 2020 analysis~\cite{Rosenberg:2022sdy}, while the Planck low-$\ell$ data use the $\texttt{Commander}$ and $\texttt{SimAll}$ likelihoods from the Planck 2018 data release~\cite{Planck:2019nip}. We refer to this combined dataset as "$\mathbf{Planck}$".
    
    \item The CMB lensing from Planck PR4~\cite{Carron:2022eyg,Carron:2022eum} is included, which is referred  as "$\mathbf{lensing}$".
    
    \item The Baryon Acoustic Oscillation (BAO) measurements from DESI DR2~\cite{DESI:2025zgx,DESI:2025zpo} include tracers of the bright galaxy samples, such as luminous red galaxies, emission line galaxies, quasars, and Lyman-$\alpha$. The corresponding measurements are summarized in Table $\text{\MakeUppercase{\romannumeral 4}}$ of Ref.~\cite{DESI:2025zgx}. We refer to this data as "$\mathbf{BAO}$".
    
    \item The compressed and foreground-marginalized likelihoods for ACT DR6 high-$\ell$ TT, TE, and EE data~\cite{ACT:2025fju,ACT:2025tim} are combined with Planck high-$\ell$ data at $\ell < 1000$ in TT and $\ell < 600$ in TE/EE from the PR3 likelihoods, as well as the low-$\ell$ Planck 2018 temperature likelihood, with the Sroll2 likelihood substituted for low-$\ell$ polarization~\cite{Planck:2019nip}. We refer to this combined dataset as "$\mathbf{Planck+ACT}$". Notably, this dataset will not be jointly analyzed with the $\mathbf{Planck}$ one to avoid reusing the Planck data.
\end{itemize}

Below we present our fitting results for different scenarios, while the full triangle plots for all cases are delegated to Appendix~\ref{Tri}. 

\subsection{Neutrino Mass Normal Ordering}\label{SecNO}
We begin by analyzing the benchmark DM-$\nu$ interacting model with normal ordering, which includes two massless and one massive flavor. For the massive neutrino, we adopt the treatment of temperature-dependent DM-$\nu$ scatterings in the Boltzmann hierarchy given in Sec.~\ref{TH}. On the other hand, the interactions between DM and massless neutrinos will be accounted for by the $T^2$-dependent cross section as in Ref.~\cite{Wilkinson:2014ksa}. The constraints on cosmological parameters in this scenario are reported in Table~\ref{table:nondeg}. It is found that the $95\%$ CL upper limit on the interaction strength is given by $u^0_{\chi-\nu}<4.21\times10^{-13}$ from the Planck dataset. By further including the CMB lensing and BAO data, the upper bound will be strengthened by a factor of 1.23 and 1.66, respectively. However, when using the Planck-ACT combined CMB data together with BAO, the constraint of the DM-neutrino interaction would be relaxed to $u^0_{\chi-\nu} < 4.06\times 10^{-13}$. In comparison with the temperature-independent case in Ref.~\cite{Mosbech:2020ahp,Wilkinson:2013kia} where the $95\%$ CL upper limit for the DM-$\nu$ coupling was $u_{\chi-\nu}\lesssim 10^{-4}$, the constraint on today's interaction parameter $u^0_{\chi-\nu}$ is strengthened by nearly nine orders of magnitude. The reason for this tighter upper bound lies in the fact that the DM-neutrino scattering cross section increases as a quadratic function of neutrino temperature when $T \gg m_\nu$, so that it could leave more imprints in the CMB and matter power spectra. 



\begin{figure}[htbp]
    \centering
    \includegraphics[width=1\textwidth]{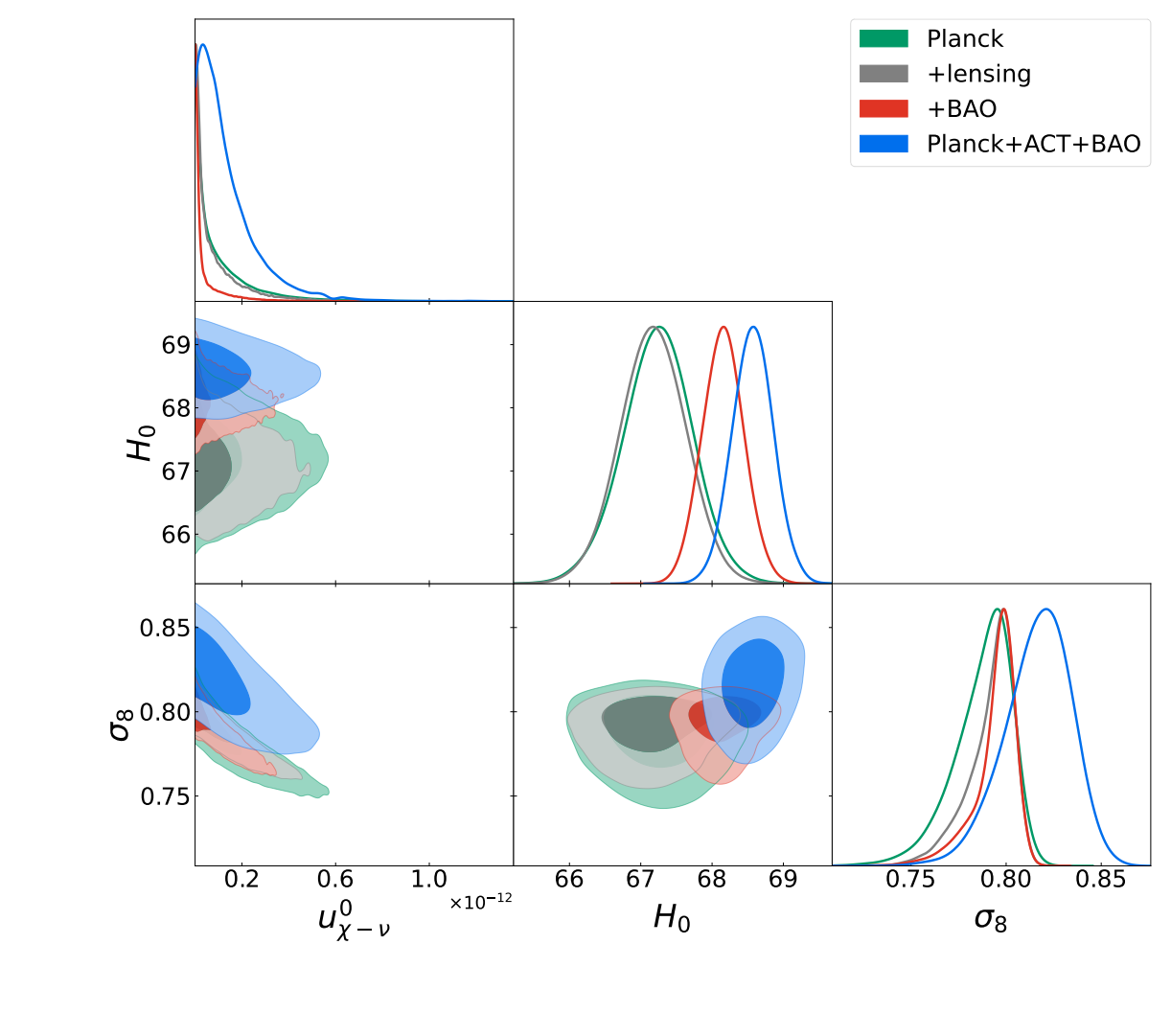}
    \caption{\textbf{Normal ordering:} One-dimensional marginalized posterior probability distributions and two-dimensional joint contours for $u^0_{\chi-\nu}$, $H_0$ [km/s/Mpc], and $\sigma_8$ obtained by analyzing different datasets listed in the legends. }
    \label{fig:three_para}
\end{figure}

Our investigation of temperature-dependent DM-$\nu$ interactions also sheds light on the long-standing $H_0$ and $\sigma_8$ problems in the literature. As we know, the measurements of the current Hubble parameter $H_0$ from the early and late universe~\cite{Planck:2018vyg,Riess:2019cxk} have exhibited more than $5\sigma$ discrepancy. Also, $\sigma_8$ quantifies the amplitude of matter density fluctuations. Its value also shows some difference between CMB observations and weak lensing measurements~\cite{Hildebrandt:2016iqg,Planck:2018vyg,DES:2021vln}. However, the latest KiDS-Legacy result~\cite{Wright:2025xka} $S_8\equiv \sigma_8\sqrt{\Omega_m/0.3}= 0.815^{+0.016}_{-0.021}$ seems to favor the CMB value, which makes the situation even more controversial (see Ref.~\cite{Pantos:2026koc} for a recent review). In our fits, $H_0$ and $\sigma_8$ are not sampled but can be derived from other parameters in the MCMC analysis, with the posterior distributions illustrated in Fig.~\ref{fig:three_para}. It is shown that no matter what datasets are used, the central values of $H_0$ are always smaller than 70, indicating that our interacting DM-neutrino model cannot solve the $H_0$ tension. In contrast, the presence of the DM-$\nu$ interaction would lead to a substantial reduction in $\sigma_8$ in the fittings. This is understandable, as the scatterings between DM and neutrinos suppress the small-scale matter power spectrum due to the DAO, causing a decrease of the matter fluctuation amplitude. Thus, our model enables to slightly alleviate the $\sigma_8$ tension without exacerbating the $H_0$ problem. Based on the fitting results in Table~~\ref{table:nondeg}, the values of $S_8$ are all predicted to be around $0.81$, which agrees with the recent KiDS-Legacy measurements very well.

\begin{table}[htb!]
\caption{\textbf{Normal ordering:} The 68\% credible intervals for cosmological parameters, except for $u^0_{\chi-\nu}$ for which only the 95\% CL upper limit is given.}
\begin{tabular}{lcccc}
\hline
Parameter & Planck & +lensing & +BAO & Planck+ACT+BAO \\
\hline
$u^0_{\chi - \nu}$ & $< 4.21\cdot 10^{-13}$ & $< 3.41\cdot 10^{-13}$ & $< 2.53\cdot 10^{-13}$ & $< 4.06\cdot 10^{-13}$ \\
$\log(10^{10} A_\mathrm{s})$ & $3.032\pm 0.016$ & $3.035\pm 0.014$ & $3.045\pm 0.014$ & $3.132\pm 0.033$ \\
$n_\mathrm{s}$ & $0.9596^{+0.0056}_{-0.0048}$ & $0.9600^{+0.0054}_{-0.0048}$ & $0.9667^{+0.0047}_{-0.0038}$ & $0.9706\pm 0.0048$ \\
$100\theta_\mathrm{s}$ & $1.04191\pm 0.00024$ & $1.04189\pm 0.00024$ & $1.04208\pm 0.00023$ & $1.04200\pm 0.00025$ \\
$\Omega_\mathrm{b} h^2$ & $0.02216\pm 0.00013$ & $0.02216\pm 0.00013$ & $0.02231\pm 0.00013$ & $0.02257\pm 0.00011$ \\
$\Omega_\mathrm{c} h^2$ & $0.1196\pm 0.0011$ & $0.1197\pm 0.0011$ & $0.11754\pm 0.00064$ & $0.11697\pm 0.00067$ \\
$\tau_\mathrm{reio}$ & $0.0513\pm 0.0076$ & $0.0526\pm 0.0070$ & $0.0588^{+0.0065}_{-0.0073}$ & $0.0573^{+0.0052}_{-0.0064}$ \\
$H_0$ [km/s/Mpc] & $67.24\pm 0.49$ & $67.17\pm 0.47$ & $68.16\pm 0.29$ & $68.57\pm 0.29$ \\
$\sigma_8$ & $0.788^{+0.018}_{-0.010}$ & $0.792^{+0.015}_{-0.0065}$ & $0.794^{+0.013}_{-0.0046}$ & $0.816^{+0.020}_{-0.015}$ \\
\hline
\end{tabular}
\label{table:nondeg}
\end{table}

\subsection{Degenerate Neutrino Masses}
We turn to the interacting DM-$\nu$ model with three degenerate neutrino masses, which has been considered for the constant scattering cross section in Ref.~\cite{Mosbech:2020ahp}. The results are presented in Table~\ref{table:deg}, from which it is seen that the $95\%$ CL upper bound in all cases are similar with $u^0_{\chi-\nu}\lesssim 8\times 10^{-11}$.   
These upper bounds are significantly stronger than the case with temperature-independent DM-$\nu$ scatterings in Ref.~\cite{Mosbech:2020ahp} for the reason already given in the first paragraph of Sec.~\ref{SecNO}. 
Moreover, the constraints on $u^0_{\chi-\nu}$ in Table~\ref{table:deg} are much weaker than those for three massless neutrino case where the interaction was constrained as $u^0_{\chi-\nu}\lesssim 10^{-14}$~\cite{DiValentino:2017oaw,Stadler:2019dii,Diacoumis:2018ezi}. This can be explained by the neutrino mass effects: when a neutrino flavor transits from relativistic to non-relativistic as the temperature drops down, the interaction of this flavor with DMs effectively turns off and thus weakens its influence on the cosmological dynamics. In sum, the above discussion indicates that incorporating the correct and realistic neutrino mass ordering is crucial to obtain the reliable constraint on the DM-$\nu$ interactions in the temperature-dependent setup.

\begin{table}[bth!]
\caption{\textbf{Degenerate masses:} The 68\% credible intervals for cosmological parameters, except for $u^0_{\chi-\nu}$ for which only the 95\% CL upper limit is presented.}
\begin{tabular}{lcccc}
\hline
Parameter & Planck & +lensing & +BAO & Planck+ACT+BAO \\
\hline
$u^0_{\chi-\nu}$ & $< 7.71\cdot 10^{-11}$ & $< 7.20\cdot 10^{-11}$ & $< 6.38\cdot 10^{-11}$ & $< 8.41\cdot 10^{-11}$ \\
$\log(10^{10} A_\mathrm{s})$ & $3.031\pm 0.016$ & $3.033\pm 0.014$ & $3.043\pm 0.014$ & $3.123\pm 0.032$ \\
$n_\mathrm{s}$ & $0.9600\pm 0.0050$ & $0.9601\pm 0.0050$ & $0.9660^{+0.0045}_{-0.0040}$ & $0.9694\pm 0.0045$ \\
$100\theta_\mathrm{s}$ & $1.04198\pm 0.00025$ & $1.04196\pm 0.00024$ & $1.04211\pm 0.00023$ & $1.04208\pm 0.00025$ \\
$\Omega_\mathrm{b} h^2$ & $0.02217\pm 0.00013$ & $0.02217\pm 0.00013$ & $0.02230\pm 0.00013$ & $0.02256\pm 0.00011$ \\
$\Omega_\mathrm{c} h^2$ & $0.1195\pm 0.0011$ & $0.1195\pm 0.0010$ & $0.11756\pm 0.00062$ & $0.11705\pm 0.00069$ \\
$\tau_\mathrm{reio}$ & $0.0513\pm 0.0077$ & $0.0520\pm 0.0070$ & $0.0580\pm 0.0068$ & $0.0572^{+0.0052}_{-0.0062}$ \\
$H_0$ [km/s/Mpc] & $67.33\pm 0.49$ & $67.30\pm 0.46$ & $68.16\pm 0.29$ & $68.57\pm 0.29$ \\
$\sigma_8$ & $0.797^{+0.011}_{-0.0086}$ & $0.7982^{+0.0092}_{-0.0059}$ & $0.7985^{+0.0086}_{-0.0055}$ & $0.823\pm 0.014$ \\
\hline
\end{tabular}
\label{table:deg}
\end{table}

\subsection{Normal ordering with varying neutrino mass}
\begin{figure}[htbp]
    \centering
    \includegraphics[width=1\textwidth]{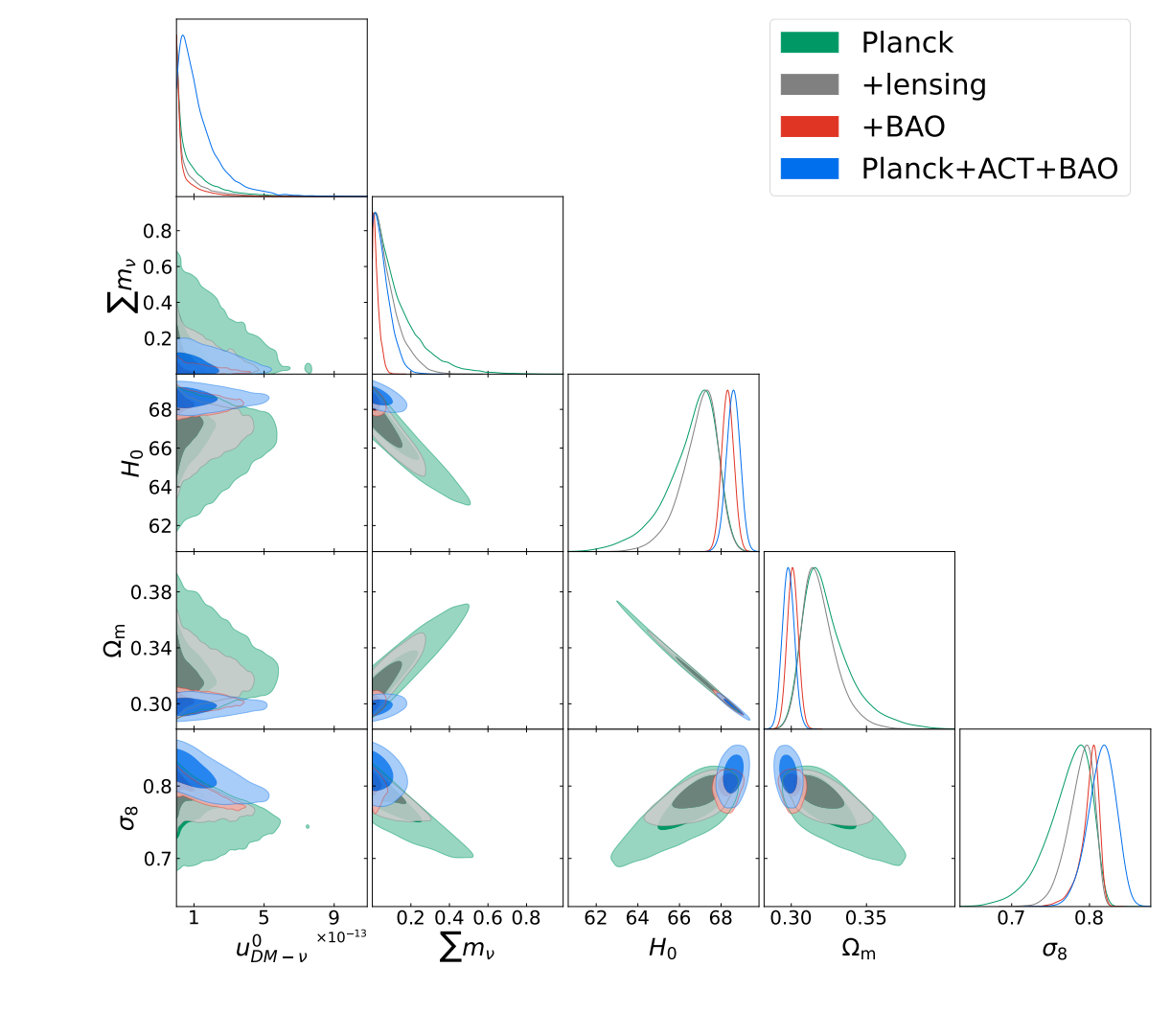}
    \caption{\textbf{Varying neutrino masses:} One-dimensional marginalized posterior probability distributions and two-dimensional joint contours for $u^0_{\chi-\nu}$, $\sum m_\nu$ [eV],  $H_0$ [km/s/Mpc], $\Omega_m$, and $\sigma_8$ obtained by analyzing different datasets listed in the legends. }
    \label{fig:vary_mass_four}
\end{figure}

In this subsection, we explore the impact of the variation of the total neutrino mass $\sum m_\nu$ on the fits of the DM-neutrino interactions. The neutrino masses are assumed to be in normal ordering. Table~\ref{table:vary_mass} summarizes the fitting results for different dataset combinations. The upper limits on the DM-neutrino interaction parameter are all around $3\times 10^{-13}$, which is similar in order as the case with fixed neutrino masses. Moreover, all fits only show upper limits on $\sum m_\nu$, with the stringent bound $\sum m_\nu \lesssim 0.102$~eV from the fit to the Planck+CMB lensing+BAO dataset. In addition, when $\sum m_\nu$ is allowed to vary, the best-fit value of the Hubble parameter $H_0$ becomes a little smaller while the range of its $95\%$ confidence interval is slightly expanded, in comparison with the fixed mass case. One can understand this result by noting that when $\sum m_\nu$ becomes larger, the value of $\Omega_m$ should increase, so that the history of cosmic expansion is altered by reducing $H_0$. The above argument can be seen more evidently from Fig.~\ref{fig:vary_mass_four}, where the contour plots for parameters $\sum m_\nu$, $H_0$, $\Omega_m$ and $\sigma_8$ are given.

\begin{table}
\caption{\textbf{Varying neutrino mass:} The 68\% credible interval for cosmological parameters, except for $u^0_{\chi-\nu}$ and $\sum m_\nu$, where only the 95\% CL upper limits are presented.}
\begin{tabular}{lcccc}
\hline
Parameter & Planck & +lensing & +BAO & Planck+ACT+BAO \\
\hline
$u^0_{\chi - \nu}$ & $< 3.98\cdot 10^{-13}$ & $< 3.13\cdot 10^{-13}$ & $< 2.11\cdot 10^{-13}$ & $< 3.74\cdot 10^{-13}$ \\
$\sum m_{\nu}$ [eV] & $< 0.428$ & $< 0.264$ & $< 0.102$ & $< 0.172$ \\
$\log(10^{10} A_\mathrm{s})$ & $3.032\pm 0.016$ & $3.039\pm 0.014$ & $3.047\pm 0.014$ & $3.148\pm 0.035$ \\
$n_\mathrm{s}$ & $0.9579^{+0.0058}_{-0.0052}$ & $0.9593^{+0.0056}_{-0.0048}$ & $0.9673^{+0.0045}_{-0.0036}$ & $0.9719\pm 0.0048$ \\
$100\theta_\mathrm{s}$ & $1.04190\pm 0.00025$ & $1.04187\pm 0.00024$ & $1.04209\pm 0.00023$ & $1.04204\pm 0.00025$ \\
$\Omega_\mathrm{b} h^2$ & $0.02211\pm 0.00015$ & $0.02212^{+0.00014}_{-0.00012}$ & $0.02232^{+0.00012}_{-0.00013}$ & $0.02259\pm 0.00011$ \\
$\Omega_\mathrm{c} h^2$ & $0.1199\pm 0.0012$ & $0.1200\pm 0.0011$ & $0.11739\pm 0.00065$ & $0.11639^{+0.00090}_{-0.00075}$ \\
$\tau_\mathrm{reio}$ & $0.0516\pm 0.0078$ & $0.0537\pm 0.0072$ & $0.0598^{+0.0066}_{-0.0074}$ & $0.0574^{+0.0053}_{-0.0063}$ \\
$H_0$ [km/s/Mpc] & $66.0^{+1.4}_{-0.69}$ & $66.37^{+0.96}_{-0.57}$ & $68.08\pm 0.30$ & $68.43\pm 0.31$ \\
$\sigma_8$ & $0.762^{+0.032}_{-0.017}$ & $0.779^{+0.020}_{-0.010}$ & $0.792^{+0.012}_{-0.0054}$ & $0.812^{+0.020}_{-0.015}$ \\
\hline
\end{tabular}
\label{table:vary_mass}
\end{table}

\subsection{Normal Neutrino Mass Ordering with a Logarithmic Flat Prior of $u^0_{\chi-\nu}$}
\begin{figure}[htbp]
    \centering
    \includegraphics[width=0.8\textwidth]{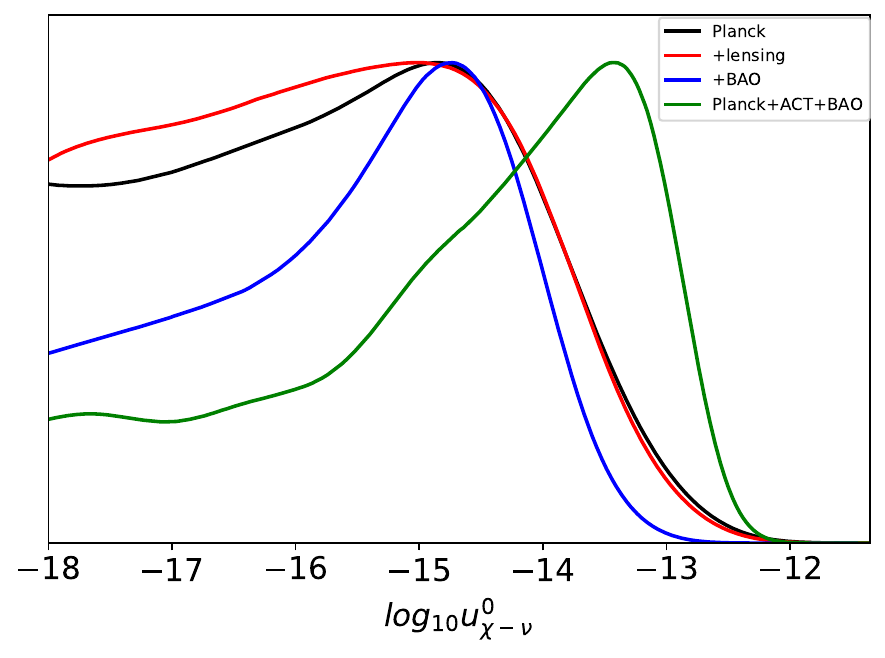}
    \caption{\textbf{Normal neutrino mass ordering with the logarithmic flat prior of $u^0_{\chi-\nu}$:} One-dimensional marginalized posterior probability distributions for $\log_{10}u^0_{\chi-\nu}$ obtained by analyzing different datasets listed in the legends. }
    \label{fig:log_u}
\end{figure}

All the aforementioned analyses adopted the linear flat prior for the DM-neutrino interaction $u^0_{\chi-\nu}$. However, some recent studies~\cite{Brax:2023rrf,Brax:2023tvn,Zu:2025lrk} have applied the logarithmic flat prior of $u^0_{\chi-\nu}$ and yielded the mild evidence for a nonzero interaction parameter. Especially, Ref.~\cite{Diacoumis:2018ezi} has explicitly shown that the final posterior distributions obtained from these two types of priors exhibited significant differences. Therefore, this subsection is devoted to investigating the effect of the logarithmic flat prior on our model fits.
The final results are presented in Table~\ref{log} and Fig.~\ref{fig:log_u}. When only the Planck temperature and polarization data are utilized, a trend for a nonzero coupling parameter is evident at $68\%$~CL as  $\log_{10}u^0_{\chi-\nu}=-15.6\pm 1.3$, while an upper bound of $\log_{10}u^0_{\chi-\nu}<-13.5$ is obtained at $95\%$ CL. Moreover, by combining the CMB lensing and BAO datasets, a clearer indication of the non-vanishing coupling with $\log_{10} u^0_{\chi-\nu}=-15.5^{+1.7}_{-2.4}$ is observed at the $95\%$ CL. By further including the small-scale CMB data from ACT, the best-fit value of $\log_{10}u^0_{\chi-\nu}$ increases to $\log_{10}u^0_{\chi-\nu}=-14.6^{+2.0}_{-3.4}$, which shows the importance to combine the large- and small-scale CMB measurements so as to enhance the sensitivity of DM interactions. 

\begin{table}
\caption{\textbf{Normal ordering with logarithmic flat priors:} The 68\% credible interval for the cosmological parameters, except for $\log_{10}u^0_{\chi-\nu}$ for which the 95\% credible interval is presented.}
\begin{tabular}{lcccc}
\hline
Parameter & Planck & +lensing & +BAO & Planck+ACT+BAO \\
\hline
$\log_{10}u^0_{\chi - \nu}$ & $-15.6\pm 1.3$ ($<-13.5$) & $-15.7\pm 1.3$ ($<-13.6$) & $-15.5^{+1.6}_{-0.91}$ ($-15.5^{+1.7}_{-2.4}$) & $-14.6^{+1.8}_{-0.67}$ ($-14.6^{+2}_{-3.4}$) \\
$\log(10^{10} A_\mathrm{s})$ & $3.034\pm 0.016$ & $3.036\pm 0.014$ & $3.045\pm 0.014$ & $3.126\pm 0.032$ \\
$n_\mathrm{s}$ & $0.9641^{+0.0044}_{-0.0040}$ & $0.9643\pm 0.0041$ & $0.9691\pm 0.0034$ & $0.9755^{+0.0049}_{-0.0040}$ \\
$100\theta_\mathrm{s}$ & $1.04188\pm 0.00024$ & $1.04187\pm 0.00024$ & $1.04205\pm 0.00023$ & $1.04195\pm 0.00024$ \\
$\Omega_\mathrm{b} h^2$ & $0.02220\pm 0.00013$ & $0.02221\pm 0.00012$ & $0.02233\pm 0.00012$ & $0.02261\pm 0.00011$ \\
$\Omega_\mathrm{c} h^2$ & $0.1194\pm 0.0011$ & $0.1194\pm 0.0010$ & $0.11746\pm 0.00062$ & $0.11688\pm 0.00068$ \\
$\tau_\mathrm{reio}$ & $0.0512\pm 0.0078$ & $0.0521\pm 0.0071$ & $0.0583\pm 0.0069$ & $0.0580^{+0.0053}_{-0.0064}$ \\
$H_0$ [km/s/Mpc] & $67.33\pm 0.49$ & $67.32\pm 0.44$ & $68.20\pm 0.28$ & $68.63\pm 0.29$ \\
$\sigma_8$ & $0.8034^{+0.0084}_{-0.0070}$ & $0.8046^{+0.0066}_{-0.0057}$ & $0.8030\pm 0.0060$ & $0.830\pm 0.015$ \\
\hline
\end{tabular}
\label{log}
\end{table}

\section{Conclusion}\label{conclusion}
We have developed a full Boltzmann hierarchy for the model with temperature-dependent scatterings between DM and neutrinos, which is induced by a dimension-six operator given in Eq.~\eqref{lagrangian}. We also devise a novel scheme to derive the fluid approximation for the modes entering the horizon deeply. In particular, we work in a realistic situation where the three neutrino masses obey the normal ordering, {\it i.e.}, we have included two massless and one massive flavors. As a result, we can study the massive neutrino effects more carefully. 
This framework has been implemented in the publicly available Boltzmann code $\texttt{CLASS}$, which is then utilized in conjunction with the MCMC sampler $\texttt{Cobaya}$ to scan and constrain the parameter space. The dataset in our fits include diverse combinations of recent cosmological observations from Planck PR4, DESI-DR2, and ACT-DR6. 

It is found that the temperature-dependent interaction can produce the DAO in the early Universe due to the pressure in the DM-neutrino fluid, which can be evident in the CMB and matter power spectra. Similar effects were reported previously in the literature for different DM-neutrino interaction scenarios. Compared with models of temperature-independent DM-$\nu$ scatterings, the constraints on today's interaction strength $u^0_{\chi-\nu}$ are stronger by seven to nine orders of magnitude, which arises since the DM-$\nu$ scattering cross section increases rapidly in the early Universe when the temperature was higher. On the other hand, our upper limits on $u^0_{\chi-\nu}$ are weaker than those in the massless neutrino scenario which predicts the $T^2$-dependent interaction. This can be understood as follows: when one neutrino flavor become non-relativistic in the late Universe, its scatterings with DM are greatly suppressed, leading to the effective decoupling. Therefore, our corresponding constraint on the DM-neutrino interaction is relaxed.  



We have payed attention to various issues related to the fits of our interacting DM model. One is the neutrino mass effects. We have considered three scenarios: normal orderings with a fixed total mass $\sum m_\nu = 0.06$~eV, fixed degenerate masses, and normal orderings with varying total mass. It turns out that the model with mass degeneracy leads to an upper bound weaker than that of the normal ordering case by nearly two orders. As for the case of varying neutrino masses, only the upper bounds on the total neutrino mass are obtained, while the limits on $u^0_{\chi-\nu}$ are similar to those with fixed masses.   Furthermore, when using logarithmic flat priors for the sampling of $u^0_{\chi-\nu}$, we observe hints of non-zero interactions at $95\%$ CL. This signal becomes more stable by combining BAO data from DESI and CMB high-multipole data from ACT. These findings indicate that future high-precision cosmological observations is important to clarify and constrain the DM-neutrino interactions. 

We have also studied the impact of the DM-neutrino interaction on the two key cosmological parameters $H_0$ and $\sigma_8$, which have been shown to be inconsistent among different observations. It is found that the existence of DM-$\nu$ scatterings in early Universe is able to suppress the matter fluctuations captured by $\sigma_8$ due to the induced DAO, but still consistent with the recent KiDS-Legacy results. On the other hand, our model cannot affect the $H_0$ tension, which indicates that its solution needs additional mechanisms.  


\begin{acknowledgments}
	\noindent This work is supported in part by the National Key Research and Development Program of China under Grant No. 2021YFC2203003 and No.~2024YFC2207204, by the National Astronomical Observatories, Chinese Academy of Sciences under Grant No. E4TG6601. 
\end{acknowledgments}

\appendix
\section{Derivation of the Collision Term and Boltzmann Hierarchy}\label{Derivation}
This appendix presents the details for deriving collision terms and the modified Boltzmann hierarchy in our model with the DM-neutrino interaction in Eq.~\eqref{lagrangian}. We shall follow the framework given in Refs.~\cite{Mosbech:2020ahp} and~\cite{Dodelson:1993xz}, with key steps summarized as follows:
\subsection{Collision Terms}\label{SubCT}
We start by considering the collision term in the Boltzmann equations for massive neutrinos, described by the following scattering integral between neutrinos and DM particles:
\begin{align}\label{collision_term}
C\left(p\right) &= \frac{1}{2E_\nu\left(\mathbf{p}\right)} \int \frac{d^3\mathbf{p'}}{(2\pi)^3 2E_\nu\left(\mathbf{p'}\right)} \frac{d^3\mathbf{q}}{(2\pi)^3 2E_\chi(\mathbf{q})} \frac{d^3\mathbf{q'}}{(2\pi)^3 2E_\chi(\mathbf{q'})}  \\
&\quad \times (2\pi)^4 |\mathcal{M}|^2 \delta^4 \left( q + p - q' - p' \right) \left[ g(\mathbf{q}') f(\mathbf{p}') (1 - f(\mathbf{p})) - g(\mathbf{q}) f(\mathbf{p}) (1 - f(\mathbf{p}')) \right]\,,\nonumber
\end{align}
where $f$ and $g$ are the distribution functions of neutrinos and DM, respectively; $\mathbf{p}$ and $\mathbf{p}^\prime$ denote the incoming and outgoing momenta for neutrinos, while $\mathbf{q}$ and $\mathbf{q}'$ denote those for DM. The delta function enforces the conservation of energy and momentum in the scattering process.  

Under the assumption that DM particles are highly non-relativistic, the DM distribution can be approximated by a delta function $g(\mathbf{q})\approx n_\chi (2\pi)^3\delta(\mathbf{q}-m_\chi \mathbf{v_\chi})$ with $\mathbf{v_\chi}$ as the DM bulk velocity, while its energy is given by $E_\chi\left(\mathbf{q}\right)\approx m_\chi + \mathbf{q}^2/(2m_\chi)$. In this way, the incoming and outgoing momenta of DM can be easily integrated over in the collision term, leaving us the following reduced form~\cite{Mosbech:2020ahp}:
\begin{align}
	C(p) &= \frac{n_\chi}{16E_\nu\!\left(\mathbf{p}\right)\,m_\chi^2} \int \frac{dp' \, d\Omega^\prime \, p'^2}{(2\pi)^3 E_\nu\left(\mathbf{p'}\right)} \frac{(2\pi)^4 \left|\mathcal{M}\right|^2}{(2\pi)^3} \nonumber \\
	&\quad \times \bigg[
	\delta\left(E_\nu\left(\mathbf{p}\right) - E_\nu\left(\mathbf{p'}\right)\right) \left[f^{(1)}\left(\mathbf{p'}\right) - f^{(1)}\left(\mathbf{p}\right)\right] \nonumber \\
	&\quad + \left(\mathbf{p} - \mathbf{p'}\right) \cdot \mathbf{v}_\chi \frac{E_\nu\left(\mathbf{p'}\right)}{p'} \frac{\partial \delta\left(E_\nu\left(\mathbf{p}\right) - E_\nu\left(\mathbf{p'}\right)\right)}{\partial p'} \left[f^{(0)}\left(p'\right) - f^{(0)}\left(p\right)\right]
	\bigg]\,,
\end{align}
where we have separated the neutrino distribution as $f(\mathbf{p}) = f^{(0)}(p) + f^{(1)}(\mathbf{p})$ with $f^{(0)}(p)$ ($f^{(1)}(\mathbf{p})$) as the background (perturbation). 




To proceed, we need to specify the form of spin averaged amplitude squared, which is given by Eq.~\eqref{scattering_amplitude_a} for our dimension-six operator of Eq.~\eqref{lagrangian}. By further assuming the DM velocity direction as the $z$-axis and defining $\mu = \hat{\mathbf{v}}_{\chi} \cdot \hat{\mathbf{p}}$ and $\mu^\prime = \hat{\mathbf{v}}_{\chi} \cdot \hat{\mathbf{p}}^\prime$, we can decompose the solid angle element as $d\Omega^\prime = d\mu^\prime d\phi^\prime$ so that the collision term can be simplified as follows
\begin{align}
	C\left(p\right) &= \frac{\sigma_{\chi-\nu}^0 n_\chi}{4\pi E_\nu\left(\mathbf{p}\right)}
	\frac{E_\nu^2\left(\mathbf{p}\right)}{E_{\nu,0}^2\left(\mathbf{p}\right)}
	\int \frac{dp'p'^2}{E_\nu\left(\mathbf{p}'\right)}\nonumber\\
	&\times\bigg[\delta\left(E_\nu\left(\mathbf{p}\right)-E_\nu\left(\mathbf{p}'\right)\right)
	\int^{1}_{-1}d\mu'\left[f^{(1)}\left(\mathbf{p'}\right) - f^{(1)}\left(\mathbf{p}\right)\right]
	\int^{2\pi}_{0}d\phi'\left(2-\frac{E_\nu\left(\mathbf{p}'\right)}{E_\nu\left(\mathbf{p}\right)}+\frac{\mathbf{p}\cdot\mathbf{p}'}{E_\nu^2\left(\mathbf{p}\right)}\right)\nonumber\\
	&+\frac{E_\nu\left(\mathbf{p}'\right)v_\chi}{p'}\frac{\partial\delta\left(E_\nu\left(\mathbf{p}\right)-E_\nu\left(\mathbf{p}'\right)\right)}{\partial p'}
	\left[f^{(0)}\left(p'\right) - f^{(0)}\left(p\right)\right]\nonumber\\
	&\times\int^{1}_{-1}d\mu'\left[\mu p-\mu'p'\right]\int^{2\pi}_{0}d\phi'\left(2-\frac{E_\nu\left(\mathbf{p}'\right)}{E_\nu\left(\mathbf{p}\right)}+\frac{\mathbf{p}\cdot\mathbf{p}'}{E_\nu^2\left(\mathbf{p}\right)}\right)
	\bigg].
\end{align}

Note that the term ${\mathbf{p}\cdot\mathbf{p}'}/{E_\nu^2\left(\mathbf{p}\right)}$ can be rewritten with the Legendre polynomials as
\begin{align}
	\frac{\mathbf{p}\cdot\mathbf{p}'}{E_\nu^2\left(\mathbf{p}\right)}=\frac{pp'}{E_\nu^2\left(\mathbf{p}\right)}P_1\left(\hat{\mathbf{p}}\cdot\hat{\mathbf{p}}'\right)\,.
\end{align}
By using the addition theorem of spherical harmonics
\begin{align}
	\int^{2\pi}_0d\phi'\,P_1\left(\hat{\mathbf{p}}\cdot\hat{\mathbf{p}}'\right)=2\pi P_1\left(\mu\right)P_1\left(\mu'\right)\,,
\end{align}
and defining the moments of the neutrino distribution function
\begin{align}\label{moments}
	\left(-i\right)^{l}f_{l}\left(p\right)=\int^{1}_{-1}\frac{d\mu}{2}P_l\left(\mu\right)\,f(p,\mu)\,,
\end{align}
the angular integrals can be performed, yielding the following collision term
\begin{align}
	C\left(p\right) &= \frac{\sigma_{\chi-\nu}^0 n_\chi}{ E_\nu\left(p\right)}
	\frac{E_\nu^2\left(p\right)}{E_{\nu,0}^2\left(p\right)}
	\int \frac{dp'p'^2}{E_\nu\left(p'\right)}\nonumber\\
	&\times\bigg[\delta\left(E_\nu\left(p\right)-E_\nu\left(p'\right)\right)
	\Bigg(\left(f^{\left(1\right)}_{0}\left(p'\right)-f^{\left(1\right)}\left(p,\mu\right)\right)\left(2-\frac{E_\nu\left(p'\right)}{E_\nu\left(p\right)}\right)-i\mu\frac{pp'}{E_\nu^2\left(p\right)}f^{\left(1\right)}_{1}\left(p'\right)\Bigg)\nonumber\\
	&+\frac{E_\nu\left(p'\right)v_\chi\mu p}{p'}\frac{\partial\delta\left(E_\nu\left(p\right)-E_\nu\left(p'\right)\right)}{\partial p'}
	\left[f^{(0)}\left(p'\right) - f^{(0)}\left(p\right)\right]\left(2-\frac{E_\nu\left(p'\right)}{E_\nu\left(p\right)}-\frac{p'^2}{3 E_\nu^2\left(p\right)}\right)\bigg]\,,
\end{align}
where the energy is only the function of the magnitude of momentum, {\it i.e.}, $E_\nu (\mathbf{p}) = E_\nu (p)$.  



Note that there are two terms in the square bracket, which will be integrated separately. Thanks to the delta function $\delta\left(E_\nu\left(p\right)-E_\nu\left(p'\right)\right)$, the momentum integral in the first term can be evaluated easily, with the result given by
\begin{align}
	&\int \frac{dp'p'^2}{E_\nu\left(p'\right)}\delta\left(E_\nu\left(p\right)-E_\nu\left(p'\right)\right)\Bigg(\left(f^{\left(1\right)}_{0}\left(p'\right)-f^{\left(1\right)}\left(p,\mu\right)\right)\left(2-\frac{E_\nu\left(p'\right)}{E_\nu\left(p\right)}\right)-i\mu\frac{pp'}{E_\nu^2\left(p\right)}f^{\left(1\right)}_{1}\left(p'\right)\Bigg)\nonumber\\
	&=p\left(f^{\left(1\right)}_{0}\left(p\right)-f^{\left(1\right)}\left(p,\mu\right)-i\frac{p^2\mu}{E_\nu^2\left(p\right)} f^{\left(1\right)}_{1}\left(p\right)\right)
\end{align}
As for the second part, we can eliminate the derivative of the delta function with the integration by parts
\begin{align}
    &\int \frac{dp'p'^2}{E_\nu\left(p'\right)}\frac{E_\nu\left(p'\right)v_\chi\mu p}{p'}\frac{\partial\delta\left(E_\nu\left(p\right)-E_\nu\left(p'\right)\right)}{\partial p'}
    \left[f^{(0)}\left(p'\right) - f^{(0)}\left(p\right)\right]\left(2-\frac{E_\nu\left(p'\right)}{E_\nu\left(p\right)}-\frac{p'^2}{3 E_\nu^2\left(p\right)}\right)\nonumber\\
    &= v_\chi\mu p\bigg(\left[p'\delta\left(E_\nu\left(p\right)-E_\nu\left(p'\right)\right)
    \left[f^{(0)}\left(p'\right) - f^{(0)}\left(p\right)\right]\left(2-\frac{E_\nu\left(p'\right)}{E_\nu\left(p\right)}-\frac{p'^2}{3 E_\nu^2\left(p\right)}\right)\right]^{p'=\infty}_{p'=0}\nonumber\\
    &-\int dp'\delta\left(E_\nu\left(p\right)-E_\nu\left(p'\right)\right)
    \left[f^{(0)}\left(p'\right) - f^{(0)}\left(p\right)\right]\left(2-\frac{E_\nu\left(p'\right)}{E_\nu\left(p\right)}-\frac{p'^2}{3 E_\nu^2\left(p\right)}\right)\nonumber\\
    &-\int dp'p'\delta\left(E_\nu\left(p\right)-E_\nu\left(p'\right)\right)
    \left[f^{(0)}\left(p'\right) - f^{(0)}\left(p\right)\right]\frac{\partial\left(2-\frac{E_\nu\left(p'\right)}{E_\nu\left(p\right)}-\frac{p'^2}{3 E_\nu^2\left(p\right)}\right)}{\partial p'}\nonumber\\ 
    &-\int dp'p'\delta\left(E_\nu\left(p\right)-E_\nu\left(p'\right)\right)
    \frac{\partial f^{(0)}\left(p'\right)}{\partial p'}\left(2-\frac{E_\nu\left(p'\right)}{E_\nu\left(p\right)}-\frac{p'^2}{3 E_\nu^2\left(p\right)}\right) \bigg)\nonumber\\
    &=-v_\chi\mu p E_\nu\left(p\right)\frac{\partial f^{(0)}\left(p\right)}{\partial p}\left(1-\frac{p^2}{3 E_\nu^2\left(p\right)}\right).
\end{align}
where the delta functions enforces $p = p^\prime$  in these integrals so that the factor $f^{(0)}(p^\prime)- f^{(0)}(p)$ vanishes. 


Combining the first and second parts gives the following form of the collision term:
\begin{align}\label{collision_term_A}
    C (p) = &\frac{\sigma_{\chi-\nu}^0 n_\chi
    pE_\nu\left(p\right)}{E_{\nu,0}^2\left(p\right)} \Bigg(f^{\left(1\right)}_{0}\left(p\right)-f^{\left(1\right)}\left(p,\mu\right)-i\frac{p^2\mu}{E_\nu^2\left(p\right)} f^{\left(1\right)}_{1}\left(p\right) \nonumber\\
    &-v_\chi\mu E_\nu\left(p\right)\frac{\partial f^{(0)}\left(p\right)}{\partial p}\left(1-\frac{p^2}{3 E_\nu^2\left(p\right)}\right)\Bigg).
\end{align}
\subsection{The Boltzmann Hierarchy for Massive Neutrinos}\label{SubBH}
Substituting the collision term in Eq.~\eqref{collision_term_A} into Eq.~\eqref{boltzmann_newtonian} gives the Boltzmann equations for neutrinos in the conformal Newtonian gauge 
\begin{align}\label{B0}
    & \frac{\partial \Psi}{\partial \tau} + i \frac{p}{E_\nu\left(p\right)} \left( \mathbf{k} \cdot \hat{\mathbf{n}} \right) \Psi + \frac{d \ln f^{(0)}\left(p\right)}{d \ln p} \left[ \dot{\phi} - i \frac{E_\nu\left(p\right)}{p} \left( \hat{\mathbf{k}} \cdot \hat{\mathbf{n}} \right) \psi \right] \\
    &=\frac{a\sigma_{\chi-\nu}^0 n_\chi
    pE_\nu\left(p\right)}{E_{\nu,0}^2\left(p\right)}\left(\Psi_{0}\left(p\right)-\Psi\left(p,\mu\right)-\frac{ip^2\mu}{E_\nu^2\left(p\right)} \Psi_{1}\left(p\right)-v_\chi\mu E_\nu\left(p\right)\frac{\partial f^{(0)}\left(p\right)}{\partial p}\left(1-\frac{p^2}{3 E_\nu^2\left(p\right)}\right)\right)\,,\nonumber
\end{align}
where $f^{(1)}(\mathbf{p}) \equiv f^{(0)}(p)\Psi(\mathbf{x}, \mathbf{p},\tau)$ according to the definition in Eq.~\eqref{perturbation}. Here $\hat{\mathbf{n}}$ is the unit vector in the direction of neutrino momentum $\mathbf{p}$. Due to the assumption that the DM fluid is irrotational, {\it i.e.}, $\nabla \times  \mathbf{v}_\chi = 0$, the spatial coordinate Fourier transform $\mathbf{k}$ is parallel to the DM velocity field $\mathbf{v}_\chi$. Thus, we have the identity $\hat{\mathbf{v}}_\chi\cdot\hat{\mathbf{n}} = \hat{\mathbf{k}}\cdot\hat{\mathbf{n}}=\mu$. 
Additionally, the extra scale factor $a$ on the right-hand side arises because we are working with the conformal time, rather than the physical one. 

In terms of Legendre polynomials, we can define the moments of the perturbation $\Psi$ as
\begin{align}
   \left(-i\right)^{l}\Psi_{l}\left(\mathbf{k},p,\tau\right)=\int^{1}_{-1}\frac{d\mu}{2}P_l\left(\mu\right)\,\Psi(\mathbf{k},\mu,p,\tau).
\label{Psi-moments}
\end{align}
Hence, the neutrino Boltzmann equation in Eq.~\eqref{B0} can be expanded so that the equations for the 0th, 1st, and $l$-th moments are given by 
\begin{align}
	\frac{\partial \Psi_0}{\partial \tau} &= \int^{1}_{-1}\frac{\mathrm{d}\mu}{2}\bigg[ 
	-i\frac{pk}{E_\nu\left(p\right)}\mu\Psi - \frac{\mathrm{d} \ln f^{(0)}\left(p\right)}{\mathrm{d} \ln p} \left[ \dot{\phi} - i \frac{E_\nu\left(p\right)}{p} \mu \psi\right] \nonumber \\   
	&+ C_\chi(p)\left[
	\Psi_{0}\left(p\right) - \Psi\left(p,\mu\right) - i\frac{p^2\mu}{E_\nu^2\left(p\right)} \Psi_{1}\left(p\right) - v_\chi\mu E_\nu\left(p\right)\frac{\partial f^{(0)}\left(p\right)}{\partial p}\left(1 - \frac{p^2}{3 E_\nu^2\left(p\right)}\right)
	\right]  
	\bigg]  \\
	-i\frac{\partial \Psi_1}{\partial \tau} &= \int^{1}_{-1}\frac{\mathrm{d}\mu}{2}\bigg[ 
	-i\frac{pk}{E_\nu\left(p\right)}\mu^2\Psi - \frac{\mathrm{d} \ln f^{(0)}\left(p\right)}{\mathrm{d} \ln p} \left[ \mu\dot{\phi} - i \frac{E_\nu\left(p\right)}{p} \mu^2 \psi\right] \nonumber \\   
	&+ C_\chi(p)\left[
	\mu\Psi_{0}\left(p\right) - \mu\Psi\left(p,\mu\right) - i\frac{p^2\mu^2}{E_\nu^2\left(p\right)} \Psi_{1}\left(p\right) - v_\chi\mu^2 E_\nu\left(p\right)\frac{\partial f^{(0)}\left(p\right)}{\partial p}\left(1 - \frac{p^2}{3 E_\nu^2\left(p\right)}\right)
	\right]  
	\bigg]\\
	(-i)^l\frac{\partial \Psi_l}{\partial \tau} &= \int^{1}_{-1}\frac{\mathrm{d}\mu}{2}\bigg[ 
	-i\frac{pk}{E_\nu\left(p\right)}P_l(\mu)\mu\Psi - \frac{\mathrm{d} \ln f^{(0)}\left(p\right)}{\mathrm{d} \ln p} \left[P_l(\mu)\dot{\phi} - i \frac{E_\nu\left(p\right)}{p} P_l(\mu)\mu \psi\right] \nonumber \\   
	&+ C_\chi(p)\bigg[
	P_l(\mu)\Psi_{0}\left(p\right) - P_l(\mu)\Psi\left(p,\mu\right) - i\frac{p^2P_l(\mu)\mu}{E_\nu^2\left(p\right)} \Psi_{1}\left(p\right)\nonumber\\ 
	&- v_\chi P_l(\mu)\mu E_\nu\left(p\right)\frac{\partial f^{(0)}\left(p\right)}{\partial p}\left(1 - \frac{p^2}{3 E_\nu^2\left(p\right)}\right)
	\bigg]  
	\Bigg],  \quad l \geq 2.
	\label{Psi-moments-1-2-l}
\end{align}
where we have used the quantity $C_\chi(p)$ defined in Eq.~\eqref{C_nudm} to simplify expressions.
By further utilizing the orthogonality of Legendre polynomials 
\begin{align}
    \int_{-1}^{1} P_l(\mu) P_k(\mu) \,d\mu = \frac{2}{2l+1} \delta_{lk}
\end{align}
we can obtain the following Boltzmann hierarchy for neutrinos
\begin{align}
    \frac{\partial \Psi_0}{\partial \tau} &= -\frac{pk}{E_\nu\left(p\right)} \Psi_1 - \dot{\phi} \frac{d \ln f^{(0)}\left(p\right)}{d \ln p}, \\
\frac{\partial \Psi_1}{\partial \tau} &= \frac{1}{3} \frac{pk}{E_\nu\left(p\right)} \left( \Psi_0 - 2\Psi_2 \right) - \frac{E_\nu\left(p\right) k}{3p} \psi \frac{d \ln f^{(0)}\left(p\right)}{d \ln p} \nonumber \\
&\quad -C_\chi\left(p\right)\left(1-\frac{p^2}{3E^2_\nu\left(p\right)}\right)\left(\Psi_1+\frac{1}{3}iv_\chi E_\nu\left(p\right)\frac{1}{f^{(0)}\left(p\right)}\frac{df^{(0)}}{dp}\right), \\
\frac{\partial \Psi_l}{\partial \tau} &= \frac{1}{2l + 1} \frac{pk}{E_\nu\left(p\right)} \left( l \Psi_{l-1} - (l+1) \Psi_{l+1}\right) - C_\chi\left(p\right)\Psi_l, \quad l \geq 2.
\end{align}

\section{Fluid Approximation}\label{FA}
The publicly available code \texttt{CLASS} uses the fluid approximation to describe the evolution of the neutrino distribution function to minimize computational cost in the regime where a given mode has entered the Hubble radius. The general idea is to truncate the hierarchy up to $l_{\text{max}} = 2$, so that we only need to solve the first three moments, corresponding to the density contrast, velocity divergence and shear, respectively. Various kinds of fluid approximations for massive neutrinos have been discussed in the literature~\cite{Hu:1998kj,Lewis:2002nc,Shoji:2010hm}. In this work, we shall use another way to derive the associated equations for massive neutrinos. 


The starting point is the Boltzmann hierarchy for neutrinos in Eqs.~\eqref{Bolt0}, \eqref{Bolt1} and \eqref{BoltL}, from which we hope to derive associate fluid equations. To this aim, note that the validity of the fluid approximation relies on the separation of scales in the dynamics. Concretely, the momentum $\mathbf{p}$ in the neutrino moments $\Psi_l (\mathbf{x}, \mathbf{p},\tau)$ characterizes fast processes which keep the local thermalization, while the Fourier transform of spatial coordinates $\mathbf{k}$ corresponds to the slow macroscopic perturbations caused by DM-neutrino scatterings. Therefore, we can average over the fast momenta $p$ first for $\Psi_l$ so as to obtain the macroscopic quantities like the energy density contrast $\delta_\nu$, velocity divergence $\theta_\nu$ and shear $\sigma_\nu$. Then we can take these integrated quantities out of the collision terms, leaving remaining integrals over the momenta as the relaxation coefficients. Since the equations for $\Psi_0$ and the associated density $\delta_\nu$ keep the same as the non-interacting case, we shall focus on the derivation of fluid equations for the velocity $\theta_\nu$ and the shear $\sigma_\nu$ in Eqs.~\eqref{NuTheta0} and \eqref{NuSigma0}, respectively. The collision terms on the right-hand side of these two equations can be approximated by
\begin{align}
	k &
	\frac{
		\int p^2 dp \, p f^{(0)}(p) \,
		C_\chi\left(p\right)
		\left( 1 - \frac{p^2}{3E^2_\nu(p)} \right)\Psi_1
	}{
		\int p^2 dp \, E_\nu(p) f^{(0)}(p) 
		+ \frac{1}{3} \int p^2 dp \, \frac{p^2}{E_\nu(p)} f^{(0)}(p)
	}\nonumber\\
	&\approx k 
	\frac{
		\int p^2 dp \, f^{(0)}(p) \,
		C_\chi\left(p\right)
		\left( 1 - \frac{p^2}{3E^2_\nu(p)} \right)
	}{\int p^2 dp \, f^{(0)}(p)}\frac{\int p^2 dp \, p f^{(0)}(p) \Psi_1}{
		\int p^2 dp \, E_\nu(p) f^{(0)}(p) 
		+ \frac{1}{3} \int p^2 dp \, \frac{p^2}{E_\nu(p)} f^{(0)}(p)
	}\nonumber\\
	&=\frac{
		\int p^2 dp \, f^{(0)}(p)\, 
		C_\chi\left(p\right)
		\left( 1 - \frac{p^2}{3E^2_\nu(p)}\right)
	}{\int p^2 dp \, f^{(0)}(p)}\theta_\nu\,,\label{ThetaC}
	\end{align}
	\begin{align}
	\frac{2}{3}&\frac{\int p^2 dp \frac{p^2}{E_\nu(p)} \,C_\chi\left(p\right) \Psi_2}
	{\int p^2 dp \, E_\nu(p) f^{(0)}(p) 
		+ \frac{1}{3} \int p^2 dp \, \frac{p^2}{E_\nu(p)} f^{(0)}(p)}\nonumber\\
	&\approx\frac{2}{3}\frac{\int p^2 dp \, C_\chi\left(p\right) }
	{\int p^2 dp \, f^{(0)}(p)}\frac{\int p^2 dp \, \frac{p^2}{E_\nu(p)} f^{(0)}(p) \Psi_2}{\int p^2 dp \, E_\nu(p) f^{(0)}(p) 
		+ \frac{1}{3} \int p^2 dp \, \frac{p^2}{E_\nu(p)} f^{(0)}(p)}\nonumber\\
	&=\frac{\int p^2 dp \, C_\chi\left(p\right)}
	{\int p^2 dp \, f^{(0)}(p)}\sigma_\nu\,, \label{SigmaC}
\end{align}
where the denominators in the last lines of above two equations come from the normalization for $\theta_\nu$ and $\sigma_\nu$. By combining other terms already existed in the non-interacting case, we can obtain the fluid equations for neutrinos in our model shown in Eqs.~\eqref{thetaNuF} and \eqref{sigmaNuF}. The momentum conservation requires us to add the corresponding collision term to the DM Euler equation as in Eq.~\eqref{thetaChiF}. Note that the coefficients before $\theta_\nu$ and $\sigma_\nu$ can be computed directly by \texttt{CLASS}~\cite{Lesgourgues:2011rh}.  By adopting this fluid approximation, we can greatly reduce the computational complexity while preserving the essential physical behavior of the system.



\begin{figure}[htbp]
    \centering
    \includegraphics[width=1\textwidth]{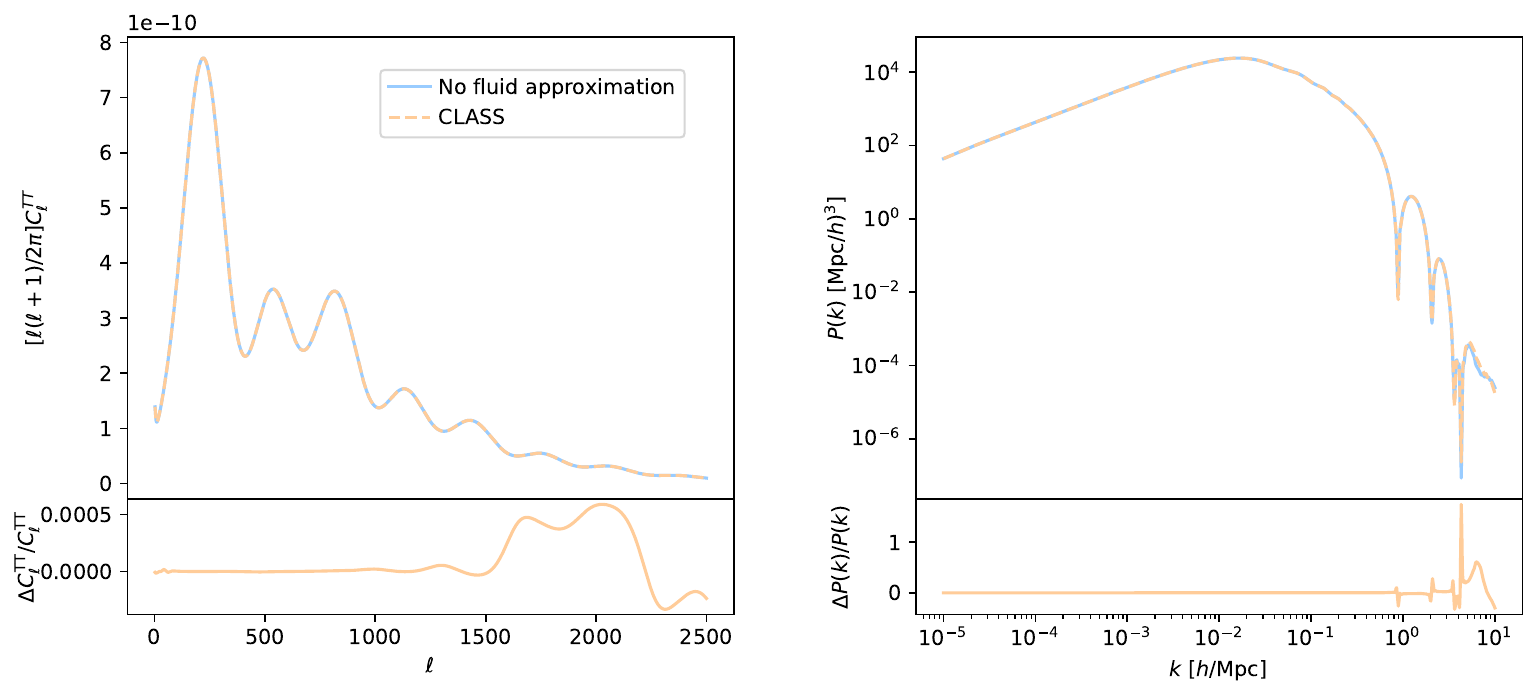}
    \caption{Comparison of the CMB TT (left panel) and matter power (right panel) spectra in our DM-$\nu$ interaction model computed either by using the full Boltzmann hierarchy or with the fluid approximation, respectively. The upper plots show the spectra, while the lower plots display the relative differences.  Cosmological parameters are set to the best-fit values obtained with the Planck dataset in Ref.~\cite{Planck:2018vyg} , while the extra interaction parameter $u^0_{\chi-\nu}$ is fixed to $10^{-10}$ in order to make the differences manifest.}
    \label{fig:appro}
\end{figure}

To demonstrate the reliability of our approximation scheme, we compute the CMB TT and matter power spectra by using both the full Boltzmann hierarchy and the above fluid approximation, with the results shown in Fig.~\ref{fig:appro}. Various cosmological parameters are set to their best-fit values obtained by the Planck Collaboration in Ref.~\cite{Planck:2018vyg}, while the interaction parameter is fixed to be $u^0_{\chi-\nu} = 10^{-10}$, in order to test the fluid approximation robustly. It is shown that the relative differences of $C^{TT}_{l}$ and $P(k)$ are smaller than $0.1\%$ in most regions, which means that these two spectra cannot be differentiated under the current experimental precision. The only exception comes in the matter power spectrum at small scales with $k \gtrsim 1\,h/\mathrm{Mpc}$, which is deep inside the nonlinear region and is not used to test our scenario with the DM-$\nu$ interaction. Therefore, the error from the fluid approximation in this region does not introduce any significant bias in our data analysis.

 \section{Triangular Plot}\label{Tri}
\begin{figure}[htbp]
     \centering
    \makebox[\textwidth][c]{\includegraphics[width=1.0\linewidth]{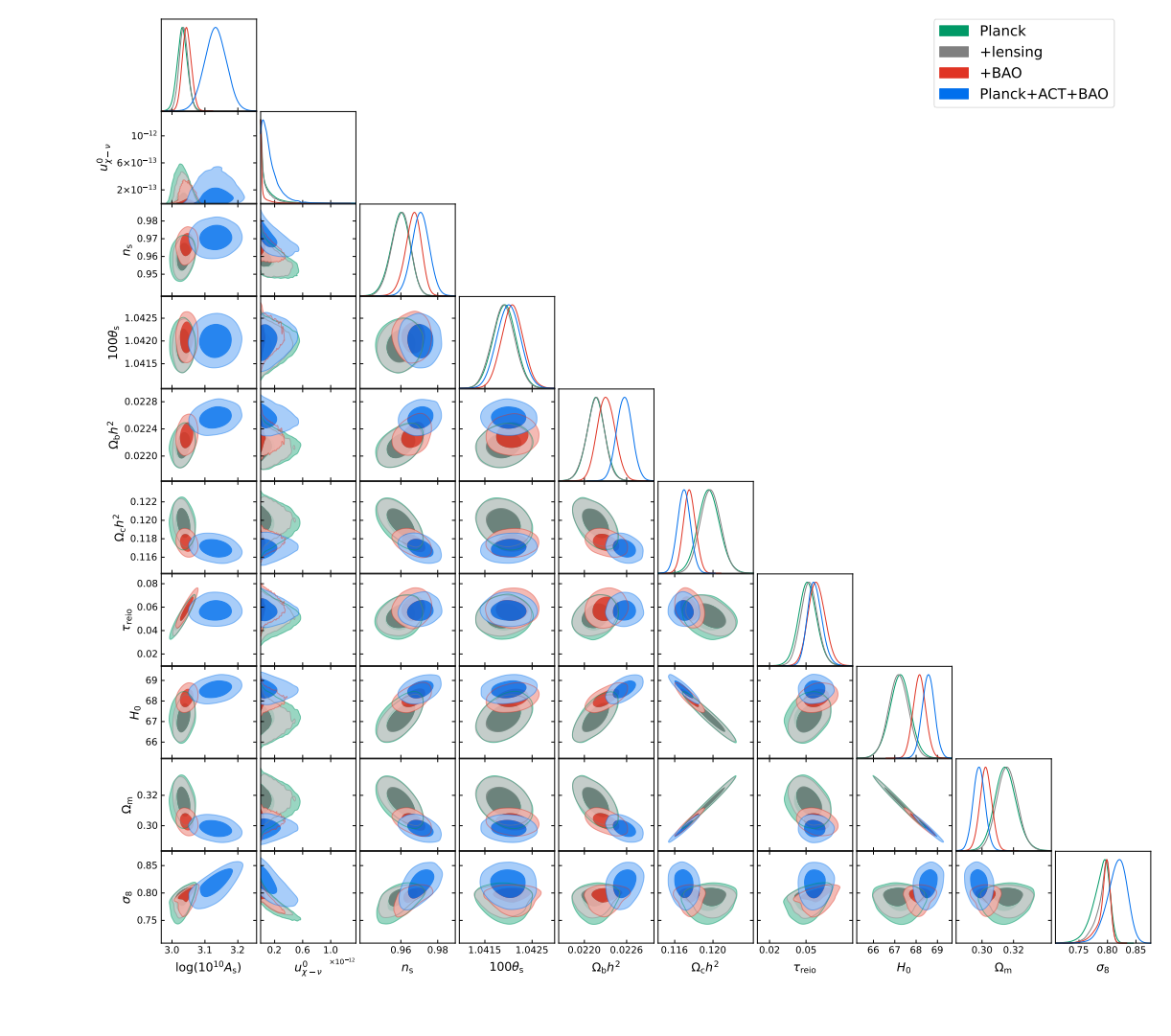}}
    \caption{One-dimensional marginalized posterior probability distributions and two-dimensional joint contours of all sampled parameters for fits with normal neutrino mass orderings obtained by analyzing different datasets, where the unit of $H_0$ is [km/s/Mpc].}
    \label{fig:nondeg}
\end{figure}

\begin{figure}[htbp]
    \centering
    \makebox[\textwidth][c]{\includegraphics[width=1.0\linewidth]{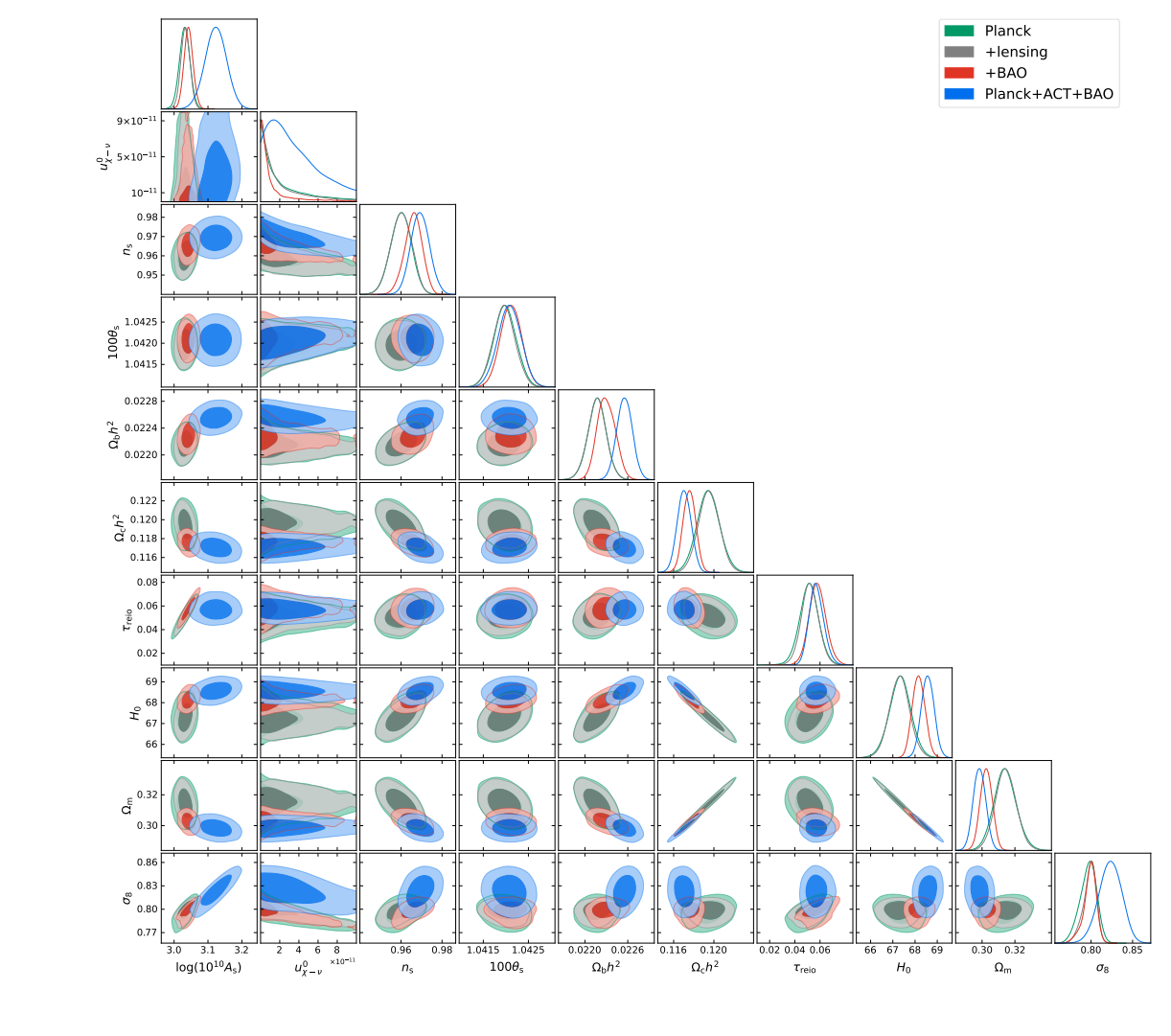}}
    \caption{The same as Fig.~\ref{fig:nondeg} except for the fits with {degenerate neutrino masses}. }
    \label{fig:deg}
\end{figure}

\begin{figure}[htbp]
    \centering
     \makebox[\textwidth][c]{\includegraphics[width=1.0\linewidth]{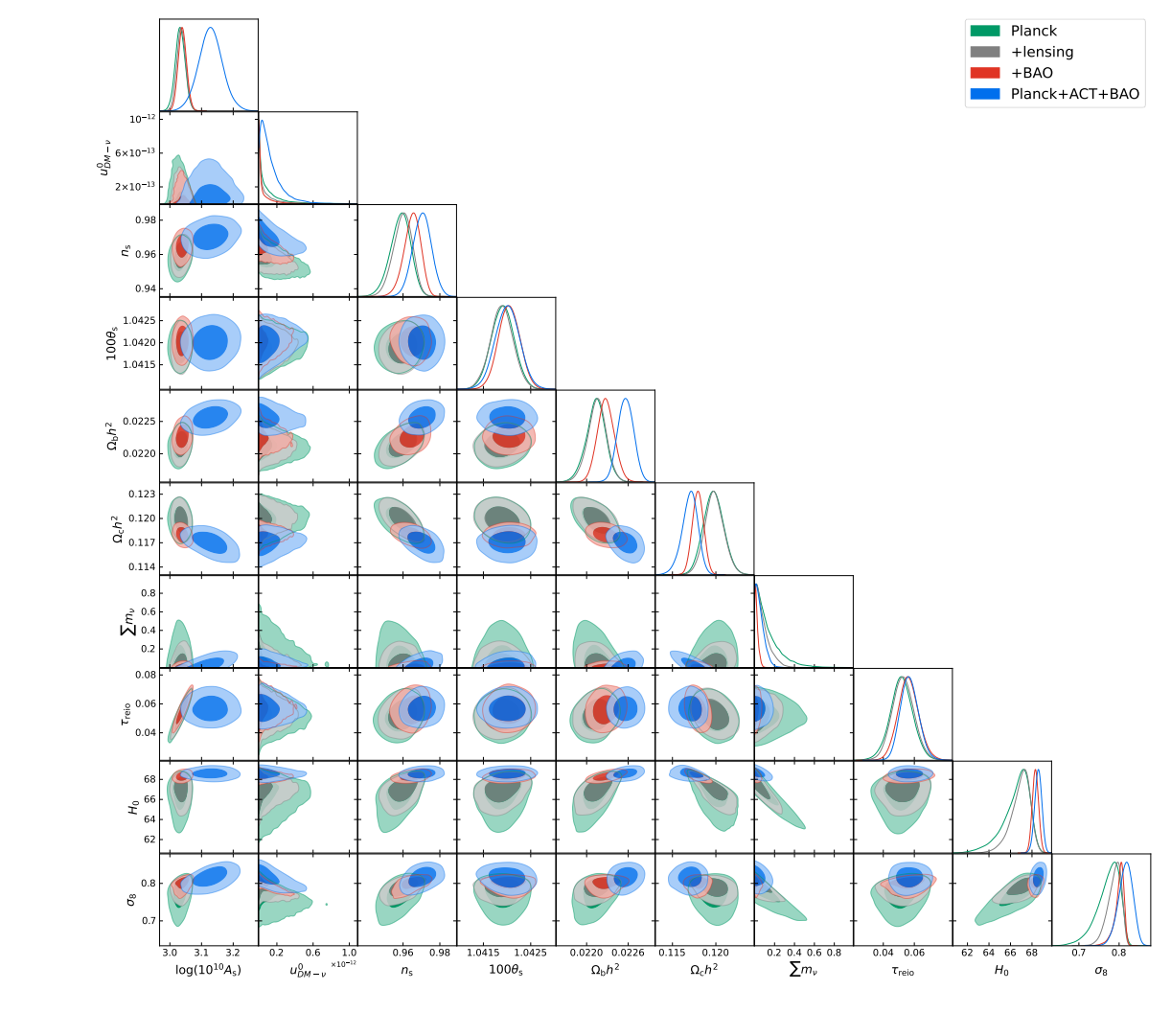}}
    \caption{The same as Fig.~\ref{fig:nondeg} except for the fits with the varying total neutrino mass, where the unit of the mass sum $\sum m_\nu$ is $[\mathrm{eV}]$. }
    \label{fig:vary_mass}
\end{figure}

\begin{figure}[htbp]
    \centering
     \makebox[\textwidth][c]{\includegraphics[width=1.0\linewidth]{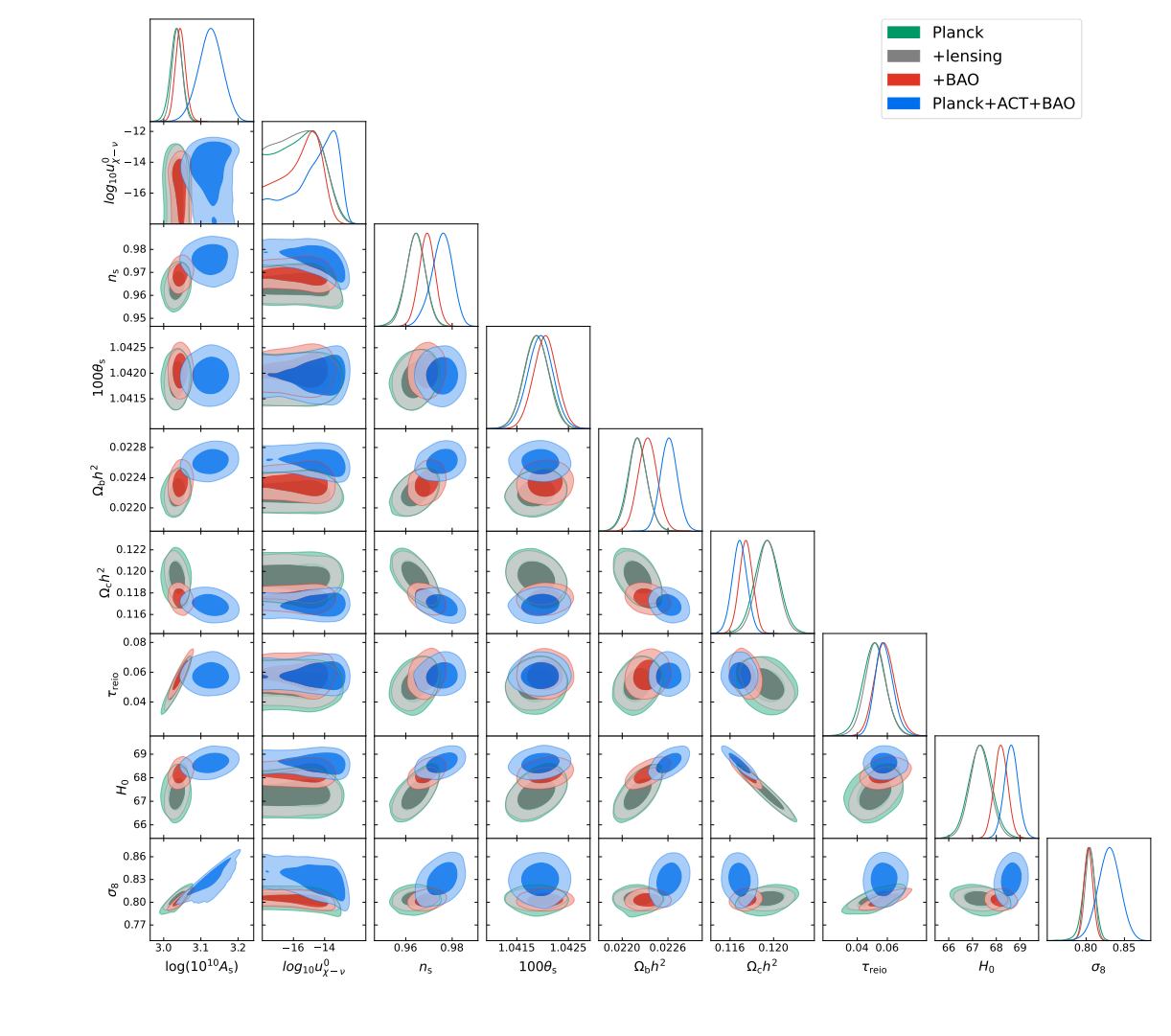}}
    \caption{The same as Fig.~\ref{fig:nondeg} except for the fits with the logarithmic flat prior for $u^0_{\chi-\nu}$.}
    \label{fig:log}
\end{figure}

 Here we show the full triangular plots for all scenarios of the DM-$\nu$ interaction in Figs.~\ref{fig:nondeg}, \ref{fig:deg}, \ref{fig:vary_mass} and \ref{fig:log}, where the green, gray, red and blue contours in each plot represent results by using the datasets of Planck, Planck+lensing, Planck+lensing+BAO, and Planck+ACT+BAO, respectively.

\bibliography{DMnu}
\end{document}